\documentclass{aa}

\usepackage{graphicx}
\usepackage{amsmath}
\usepackage{txfonts}
\usepackage{booktabs}

\usepackage[dvipsnames]{xcolor}

\begin{document}

\title{Radiation, Thermochemistry, and Stellar Feedback in DISPATCH}
\titlerunning{Radiation and Feedback in DISPATCH}

\authorrunning{Lake et al.}

\author{
William Lake\inst{1}\corrauth{williamlake4@gmail.com}
\and
Paolo Padoan\inst{1,2}\email{ppadoan@icc.ub.edu}
\and
{\AA}ke Nordlund\inst{3}\email{aake@nbi.ku.dk}
\and
Troels Haugb{\o}lle\inst{3}\email{haugboel@nbi.ku.dk}
}

\institute{
Department of Physics and Astronomy, Dartmouth College,
6127 Wilder Laboratory, Hanover, NH 03755, USA
\and
Institut de Ci\`encies del Cosmos (ICCUB),
Universitat de Barcelona, Mart\'i i Franqu\`es 1,
E-08028 Barcelona, Spain
\and
Niels Bohr Institute, University of Copenhagen,
Jagtvej 155A, DK-2200 Copenhagen N, Denmark
}

\date{}

\abstract
{
The growth of massive stars depends on whether their accretion flows survive radiation and outflow feedback from the forming stellar population.
}
{
We develop and validate a numerical framework for investigating how radiative and mechanical feedback regulate accretion onto individual stars within turbulent star-forming clouds.
}
{
We present an extension of DISPATCH that couples directional
stellar radiation, diffuse infrared transport, and thermochemistry
to self-gravitating magnetohydrodynamics, individual stellar
accretion, and protostellar jets. This AMR-based method uses DISPATCH's local timestepping, allowing stellar environments and their larger scale cluster environments to step on separate timescales, and providing the framework needed to scale to long-duration simulations of GMCs. Direct radiation is transported in multiple bands, including both ionizing and nonionizing radiation, and uses short characteristics to maintain source directions and shadowing. Diffuse radiation is implemented for IR reprocessed light, with M1 and stationary ray options for transport. The thermochemical model includes atomic and molecular hydrogen and helium chemistry, and metal cooling and dust. It also includes a composition-dependent equation of state. Stellar masses grow through accretion of gas rather than a prescribed IMF, allowing stellar masses to respond to the evolving gas supply and feedback.
}
{
In tests, multiple-source illumination retains geometric shadows while diffuse reprocessing heats shielded material, and tests of jets, ionization fronts, and radiation pressure demonstrate expected behavior and convergence. Ionization fronts also cross asynchronous refinement boundaries with late-time propagation speeds within a few percent of analytic predictions.
}
{
Together, these developments provide a framework for determining how radiative and mechanical feedback modify the duration and efficiency of accretion from turbulent clouds, connecting the gas supply to individual stars with the high-mass stellar initial mass function.
}


\maketitle
\nolinenumbers

\section{Introduction}
\label{sec:introduction}

The high-mass end of the stellar initial mass function (IMF) determines how strongly a young stellar cluster irradiates and enriches its surrounding gas. But the processes that determine the shape of this IMF, and its universality across environments, remain uncertain \citep{mckee_theory_2007,Grudic+22}. Explaining these processes requires a full understanding of the coupling between young stars and their gas: how long the gas continues to feed them, and how heavily they feed, as well as what process terminates their accretion. A central question in this problem is whether or not massive stellar growth ends primarily because the surrounding cloud ceases to supply gas or because feedback prevents that gas from reaching the star.

The inertial-inflow model \citep{Padoan+20} posits that large-scale, long-duration accretion flows formed via supersonic turbulence may dominate the accreted mass of high-mass stars, rather than 
the rapid collapse of preassembled massive stellar cores \citep{McKeeTan02,McKeeTan03}.
In this model, massive stars continue to accrete as long as they are supplied by large-scale turbulent inflows, potentially for several Myr, a timescale comparable to their main-sequence lifetimes, which has important implications for the duration of the embedded phase \citep{PadoanGieles2026} and the location of the stellar birthline in the Hertzsprung--Russell diagram \citep{PadoanShipkoff2026}.
The model predicts that massive-star formation timescales depend on the environment, becoming shorter at higher gas surface densities, and the maximum stellar mass is proportional to the parent cloud mass \citep{Padoan+20,PadoanGieles2026}. In the massive, dense clouds where globular clusters form, these scalings allow extremely massive stars to assemble within $\sim1$--$2$\,Myr \citep{Gieles+25}.

An important limitation of the inertial-inflow model is that the simulations on which it is based neglect stellar feedback,
which may erode turbulent accretion flows \citep{Dale+05} and reduce both the rates and duration of massive-star growth.
However, the STARFORGE simulations, which include radiative and mechanical feedback, also find prolonged accretion, with formation times increasing with final stellar mass \citep{Grudic+22,Wainer+26}. These results suggest that radiative feedback does not necessarily prevent massive stars from growing over several Myr, although it can play an important role in terminating accretion. A more significant quantitative discrepancy is that STARFORGE finds maximum accretion rates approximately an order of magnitude higher than predicted by the inertial-inflow model \citep{Grudic+22}. This challenges the predicted proportionality between maximum stellar mass and parent cloud mass. Nevertheless, the STARFORGE simulations yielding these higher accretion rates follow the collapse of isolated clouds initialized with a random velocity field, without sustained turbulent driving. Furthermore, the initial conditions are not evolved for several turbulent dynamical times to establish self-consistent density, velocity, and magnetic-field fluctuations and allow the turbulent amplification of the rms magnetic field \citep{Federrath+11,Beattie+23}.

Understanding the full assembly and evolution of the gas that feeds these massive stars is a challenge: one must simultaneously capture the large-scale processes of turbulent gas dynamics and cloud collapse and the small-scale processes of fragmentation, stellar accretion, feedback, and stellar dynamics. Existing works have built towards a deeper understanding of each process, and towards unification of many processes into increasingly powerful simulations with better-understood physics. Simulations of supersonic turbulence have developed our understanding of the connection between cloud structure, fragmentation, and the gas supply to stars \citep{Larson81,Padoan01,Padoan02,MacLowKlessen04,mckee_theory_2007, Haugbolle+18,Padoan+20}. Radiation-hydrodynamic (RHD) simulations have shown how radiative feedback can suppress fragmentation, as well as eroding accretion flows and dispersing star-forming clouds \citep{Offner+09,Dale+05,Dale+12,Grudic+22,Menon+25}. Meanwhile, simulations of jet outflows have found that they are capable of regulating individual star formation and influencing aspects of the IMF \citep{Cunningham+11,Federrath+14,Guszejnov+21}, as well as acting in concert with radiation to permit escape via low-density polar regions \citep{RosenKrumholz20}. These developments have been accompanied by advances in our understanding and treatment of cooling, dust, and nonequilibrium chemistry in star formation at a wider range of densities, metallicities, and radiation environments \citep{Omukai+05,Schneider+06,GloverAbel08,GloverClark12,
Grassi+14,Richings+14a,Richings+14b,Baczynski+15,Smith+17,
PloeckingerSchaye20,Lake+22,Hopkins+23,Kim+23}. State-of-the-art codes are beginning to unify many or all of these processes at increasingly large dynamic ranges and timescales: STARFORGE follows individual star formation, accretion, dynamics, and feedback on the scales of giant molecular clouds (GMCs) \citep{Grudic+20, Grudic+22}, enabling studies of the origins of the IMF, of primordial star formation, and of obscuration of protostars \citep{Guszejnov+22, Lake+25, Wainer+26, Meziani+26}. TORCH has followed gas and stellar dynamics, evolution, and feedback in simulations of massive star formation, including both radiation and outflows and simulating at the scale of globular clusters \citep{Wall+20, Polak+24, Appel+25}. Further, ORION simulations of collapsing cores with full radiation-magnetohydrodynamic (RMHD) treatments have shown how magnetic fields and protostellar outflows regulate massive-star growth and open cavities through which radiation can escape \citep{RosenKrumholz20}, and VETTAM's studies in metal-free gas have applied many of these processes to high-resolution simulations of primordial star formation, extending our understanding to a regime that is near impossible to directly observe \citep{Menon+22, ShardaMenon25}.

In our work, we focus on the coupling between radiative feedback and accretion flows that feed high-mass stars, to better understand the lifetimes of massive star growth. This requires a model capable of self-consistently resolving the IMF through individual stellar formation, growth, and evolution. The simulation must also resolve stellar dynamics, follow nonequilibrium chemistry and cooling consistently with radiative transport, and follow processes most relevant to the escape of radiative feedback in young stars: namely, protostellar jets. It must also include a mutually consistent turbulence, density, and magnetic field distributions: because the turbulence controls the properties of star-forming clumps and the magnetic field provides support that can impact the IMF, it is important to at least drive turbulence on large scales to generate initial conditions for isolated cloud runs \citep[and ultimately it is better to simulate larger boxes with self-consistent turbulence][]{Padoan+16, Padoan+17, Haugbolle+18, Lane+22}. None of these models cited above meet all these requirements: specifically, none have ray-traced radiation coupled to thermochemistry, MHD, and individual star formation and accretion (not sampled from an IMF) for Myr timescales. Ray tracing, while expensive, is important for understanding the coupling of radiation and accretion flows that may set the accretion lifetime in the inertial-inflow model: in dense clusters, where multiple bright sources illuminate outflow cavities and dense filaments, moment based methods such as the ${\rm M}_1$ closure can underestimate the impacts of direct radiative feedback \citep{Kim+17, Grudic+22}. 

Therefore, in this work we extend DISPATCH \citep{Nordlund+18} to accomodate all of these properties. DISPATCH is a modular framework for astrophysical problems, which uses adaptive mesh refinement (AMR) and asynchronous local timestepping to break problems down into parallel tasks for a dispatcher. The framework divides the box into patches, which are locally advanced according to their own evolutionary timescales, enabling highly efficient computation targeted at primarily the most rapidly evolving regions of the simulation. The architecture of DISPATCH revolves around modules, each of which solves a particular aspect of an astrophysical problem (for example, there is an MHD module, and a supernova module). DISPATCH has been used across a range of astrophysical environments, from pebble accretion and the structure of planetary envelopes \citep{Popovas+18, Popovas+19}, to stellar atmospheres and solar convection \citep{Popovas+22,Eitner+24,Hoppe+26}, to supersonic turbulence \citep{Pan+19}. This flexibility makes it an ideal platform to extend to the wide-dynamic-range problem of star cluster formation and evolution. Here, we extend DISPATCH with $5$ new modules, and associated changes to existing modules such as the MHD solver: a cooling module, a chemistry module, a point-source radiation module, an equation of state, and a protostellar jet module. In the present paper, we present and test these models, showing their capability and applicability for understanding star formation in cloud environments.

This paper is organized as follows:  in Section~\ref{sec:numerical-method}, we discuss the implementation of radiation and its coupling to the gas. In Section~\ref{sec:physical-model}, we discuss the thermochemistry, star formation, and outflow models in DISPATCH, as well as our dynamical framework. In Section~\ref{sec:validation}, we show a number of proof-of-concept simulations that test and validate our code. In Section~\ref{sec:discussion}, we compare the strengths and limitations of our code to other methods in the literature, as well as discussing avenues for future improvement. In Section~\ref{sec:conclusions}, we conclude with a summary of the method and usage opportunities.

\section{Radiation Transport and Coupled Evolution}
\label{sec:numerical-method}

We separate radiation into two components, a point-source field emitted directly from stars, and a diffuse field emitted by dust and gas. Source directions are retained explicitly in the direct field, which allows illumination of accreting flows and outflow cavities from multiple directions and geometries at once.

\subsection{Direct Stellar Radiation}
\label{sec:direct-radiation}

\begin{table}[t]
\caption{Energy intervals of the direct stellar radiation groups.}
\label{table:radfields}
\centering
\begin{tabular}{lc}
\toprule
Group & Photon energy (eV) \\
\midrule
He II-ionizing & $54.4$--$500$   \\
He I-ionizing  & $24.6$--$54.4$ \\
H-ionizing     & $13.6$--$24.6$ \\
Dissociating   & $11.2$--$13.6$    \\
FUV            & $6$--$11.2$    \\
NUV            & $3.4$--$6$     \\
Optical/NIR            & $0.41$--$3.4$     \\
IR             & $<0.41$       \\
\bottomrule
\end{tabular}

\end{table}

The direct field has $8$ spectral bands, $4$ ionizing or dissociating and $4$ non-ionizing. These are summarized in Table~\ref{table:radfields}. Within a spectral band, luminosities, photon energies, atomic cross-sections, and photoheating energies are averaged. For each spectral band $g$, we denote the radial energy flux
from a point source by $F_g(r)$ and define $\widetilde{F}_g=r^2F_g$, removing geometric dilution from the transport equation. The attenuation of direct stellar radiation is then described by
\begin{equation}
\frac{d\widetilde{F}_g}{dr}=-\alpha_g\widetilde{F}_g,
\qquad
\widetilde{F}_g(0)=\frac{L_g}{4\pi},
\end{equation}
where $\alpha_g$ is the absorption coefficient per unit length, including the relevant H\,I, He\,I, He\,II, and dust absorption,
and $L_g$ is the source luminosity in spectral band $g$.


We implement a short-characteristics ray tracing method, which steps radially outward through cells away from the source. For ionizing radiation, we adapt C2-ray \citep{Mellema+06}, which computes time-averaged ionized fractions and optical depths self-consistently across multiple sources. This allows the radiation to respond to changes in ionization state throughout the trace. Recombination uses the on-the-spot approximation. The radiation update is patch-local, and patches communicate changes in radiation when they exchange neighbouring face-based fluxes asynchronously.

Dust absorption uses band means dependent on the source temperature for each ray, coming from the $R_V=5.5$ model of \citet{Draine03} (see Section~\ref{sec:dust} for more details). The final point-source fields supply photoionization, photoheating, grain photoelectric heating, H$_2$ dissociation, and H$^-$ photodetachment rates to the thermochemical model. 

Lyman--Werner (LW) photodissociation of H$_2$ includes self-shielding using the prescription of \citet{DraineBertoldi96}, with a fixed Doppler parameter $b=3\,\mathrm{km\,s^{-1}}$. The shielding factor is normalized to unity at zero H$_2$ column density. The dissociation rate is
\begin{equation}
 k_{\rm LW}
 =c\,\sigma_{\rm LW,eff}
 \sum_a n_{\gamma,{\rm LW},a}^{\rm dust}
 f_{\rm sh}(N_{{\rm H}_2,a},b),
 \label{eq:lw-dissociation}
\end{equation}
where $n_{\gamma,{\rm LW},a}^{\rm dust}$ is the dust-attenuated,
cell-averaged LW photon density from source $a$. This treatment approximates molecular line absorption, but it does not resolve individual lines or spatially varying gas velocities.

\subsection{Diffuse Radiation}
\label{sec:diffuse-radiation}

We implement two options for diffuse IR radiation, dependent on the needs of the problem. Our fiducial diffuse IR field is handled via M1-closure-based evolution of a color-temperature-based radiation energy density ${\rm E}_{\rm IR}$ and flux $\boldsymbol F_{\rm IR}$ \citep{Levermore84}. The closure uses the reduced flux $f=|\boldsymbol F_{\rm IR}|/(cE_{\rm IR})$ and approaches $\boldsymbol P_{\rm IR}=E_{\rm IR}\boldsymbol I/3$ in the isotropic limit. Writing $\epsilon=\widehat c/c$, where $\widehat c$ is the reduced speed of light, the infrared moment equations are
\begin{align}
 \frac{\partial E_{\rm IR}}{\partial t}
 +\epsilon\boldsymbol\nabla\cdot\boldsymbol F_{\rm IR}
 &=\epsilon\left(\mathcal L_{\rm IR}+\mathcal G+\mathcal H_{\rm d}\right),\\
 \frac{\partial\boldsymbol F_{\rm IR}}{\partial t}
 +c\widehat c\,\boldsymbol\nabla\cdot\boldsymbol P_{\rm IR}
 &=-\widehat c\alpha_R\boldsymbol F_{\rm IR}.
 \label{eq:ir-moments}
\end{align}
Here $\mathcal G$ is the gas-to-dust energy transfer, $\mathcal L_{\rm IR}$ is the gas emissivity assigned to the infrared field, and $\alpha_R$ is the Rosseland transport extinction coefficient per unit length. $\mathcal H_{\rm d}$ is the stellar power (and, when enabled, the imposed interstellar radiation field) absorbed by grains, minus photoelectric energy transferred to the gas, plus energy transferred from the gas by grain recombination. See \ref{sec:dust} for a full discussion of the dust-radiation coupling treatment. 

We characterize the infrared radiation spectrum by a color temperature $T_{\rm col}$ and transport the auxiliary variable $\mathcal C_{\rm IR}=E_{\rm IR}/T_{\rm col}$ alongside the radiation energy density $E_{\rm IR}$, so that geometric dilution does not change $T_{\rm col}$.
Dust emission adds its emitted energy divided by $T_{\rm d}$ to $\mathcal C_{\rm IR}$; gas emission assigned to the infrared field uses the gas temperature. Absorption removes equal fractions of $E_{\rm IR}$ and $\mathcal C_{\rm IR}$, and dilution preserves their ratio. For M1, the Planck absorption and Rosseland transport means
are weighted at $T_{\rm col}$, whereas the Planck emission
mean is weighted at $T_{\rm d}$. Grain phase and survival
depend on the local dust state. 

Our alternative implementation of diffuse IR transport uses
stationary angular short characteristics within DISPATCH
\citep{Nordlund+18}. Intensities and their color counterparts
are transported along a discrete angular quadrature, from
which the radiation energy density and flux are obtained.
This retains directional information without an M1 closure.
To recover Planck weighting in optically thin material and
Rosseland weighting in the diffusion limit, we interpolate
between the two means,
\begin{align}
\kappa_{\rm eff}(T_{\rm w})
 &=w(T_{\rm w})\kappa_{\rm P}(T_{\rm w})
  +[1-w(T_{\rm w})]\kappa_{\rm R}(T_{\rm w}),\\
w(T_{\rm w})
 &=\exp[-2\Sigma\kappa_{\rm R}(T_{\rm w})],
\label{eq:ray-opacity-blend}
\end{align}
following the interpolation used by \citet{Eitner+24}.
Here $T_{\rm w}$ is the spectral weighting temperature;
the opacity means are expressed per unit gas mass and
include the adopted dust abundance and local grain state.

For collapsing gas, we estimate the mass column locally as
\begin{align}
\Sigma&=\rho\min\left(
  \frac{\rho}{|\boldsymbol{\nabla}\rho|},
  \frac{\lambda_{\rm J}}{2}\right),\\
\lambda_{\rm J}
 &=\left(\frac{\pi c_{\rm iso}^{2}}{G\rho}\right)^{1/2},
\label{eq:ray-opacity-column}
\end{align}
where $c_{\rm iso}^{2}=P/\rho$ is evaluated using the local
composition at the gas temperature. 
Unlike the hydrostatic column used by \citet{Eitner+24},
this estimate uses the local density structure and
self-gravitating pressure scale.

Absorption uses
$\alpha_{\rm abs}=\rho\kappa_{\rm eff}(T_{\rm col})$,
while emission uses
$\alpha_{\rm em}=\rho\kappa_{\rm eff}(T_{\rm d})$.
The column is held fixed during each local material solve,
while the emission coefficient responds to the trial dust
temperature. Applying the same interpolation to absorption
and emission preserves their equality in local thermodynamic
equilibrium and recovers Rosseland transport in the optically
thick LTE limit. The interpolation requires no additional
frequency groups or angular sweeps. The angular quadrature
is specified for each simulation run.
 
For both cases, the diffuse field includes H$_2$ line emission and the low-temperature fine-structure and molecular-metal cooling terms.

\subsection{Momentum Coupling and Local Integration}
\label{sec:radiation-matter}

When radiation pressure is enabled, radiation couples to the gas along the source direction, and diffuse radiation transfers momentum via the dust transport opacity:

\begin{equation}
 \Delta\boldsymbol p_g=
 \frac{\Delta E_{{\rm abs},g}}{c}\widehat{\boldsymbol r},
 \qquad
 \boldsymbol f_{\rm IR}=
 \frac{\alpha_{\rm mom}}{c}\boldsymbol F_{\rm IR},
 \label{eq:radiation-momentum}
\end{equation}
where the first expression controls direct absorption and the second expression is the volumetric force from IR radiation. Here $\Delta E_{{\rm abs},g}$ is the energy absorbed from direct radiation 
in spectral band
$g$, and $\boldsymbol f_{\rm IR}$ is the diffuse force per unit volume. For M1, $\alpha_{\rm mom}=\alpha_R$, and the radiation impulse is accumulated from the implicit flux damping over the radiation substeps. In the static diffusion limit, the diffuse force approaches $-\boldsymbol\nabla(E_{\rm IR}/3)$. For stationary angular transport, $\alpha_{\rm mom}=\rho\kappa_{\rm eff}(T_{\rm col})$, using the same coefficient as the angular sweep. The gas receives both direct and diffuse contributions.

Radiation follows the MHD update within a timestep, with boundary data supplied by the neighbouring patch exchange. The full chemical network is advanced once per timestep, with its associated heating and cooling. M1 transport and gas-dust-radiation energy exchange are subcycled, holding the chemical abundances fixed and retaining the corresponding line and grain energy contributions without repeating their gas-energy changes (ray transport simply limits the patch timestep via its radiative heating and cooling, with no subcycling). 

M1 IR transport uses an HLL flux with monotonized-central piecewise-linear reconstruction, followed by implicit local material exchange and flux damping. An optical-depth-dependent correction suppresses excessive numerical diffusion in opaque cells and is limited consistently with the material update.  This separation of explicit transport and implicit local sources follows the approach discussed by \citet{Chu+19,He+24}; our implementation uses one transport/source stage per substep and is first order in time.

Our M1 substeps follow 
\begin{equation}
 \widehat c\,\delta t_{\rm IR}
 \sum_{i=1}^{3}\frac{1}{\Delta x_i}\le C_{\rm IR},
 \label{eq:radiation-courant}
\end{equation}
with fiducial values $\widehat c=30$ km s$^{-1}$ and $C_{\rm IR}=0.4$ \citep[fiducial values following the general arguments of ][]{Geen+15, Grudic+20}. This reduced speed of light only applies to the diffuse field; our point-source field has no such constraint. The adequacy of $\widehat c$ depends on radiation transport and thermal relaxation relative to material evolution, including diffusion through opaque structures, and, particularly in rapid flows, $\widehat c$ may be raised depending on the problem at hand.

\section{Physical Model}
\label{sec:physical-model}

\subsection{Dynamical Framework}
\label{sec:framework}

As mentioned in Section~\ref{sec:introduction}, our new star formation method is implemented within DISPATCH \citep{Nordlund+18}. Within DISPATCH, gas is represented by Cartesian patches within an AMR framework with local timestepping. Each patch advances on its own evolutionary timescale, using neighbouring patches to supply boundary conditions at a given time. In the high dynamic range problem of a star-forming cloud, this permits much of the computational effort to be spent on collapsing cores and feedback-driven shocks, rather than turbulent, diffuse gas. 

We use the RAMSES-derived Godunov MHD solver available within DISPATCH, which uses an HLLD Riemann solver and uses constrained transport for the magnetic field. Chemical species are transported using passive scalar transport and evolved using a local reaction network, and thermodynamic quantities are derived from the composition-dependent equation of state described below. The sound speed is evaluated locally, and does not assume a uniform adiabatic index throughout the gas. 

We include gas self-gravity and gas-sink gravitational interactions. Sink-sink gravity uses unsoftened Newtonian forces, while gas-sink gravity is treated at finite resolution separately. Each sink represents an individual accreting star; stellar masses are, therefore, determined by the gas supply and the accretion prescription described below in Section~\ref{sec:stellar-sources}. 

Turbulence is driven by forcing the gas on scales comparable to the box size.
Specifically, in our star formation simulations, we drive solenoidal turbulence using Fourier-space forcing at wavenumbers $1 \leq kL/(2\pi) \leq 2$, where $L$ is the box size.
Turbulent initial conditions are generated by evolving initially uniform, magnetized gas under this forcing for several dynamical times, allowing correlated density, velocity, and magnetic-field fluctuations to develop self-consistently.
Driving continues during collapse and star formation, which leads to a sustained input of turbulence from larger scales. 

\subsection{Chemical Network and Equation of State}
\label{sec:chemistry-eos}

We follow nonequilibrium states of hydrogen (${\rm H}$ and ${\rm H}^+$), helium (${\rm He}$, ${\rm He}^+$, and ${\rm He}^{++}$), and molecular hydrogen, in a reduced chemical network. We adopt global hydrogen and helium fractions X~$=0.76$ and Y~$=0.24$. The atomic network includes collisional ionization, photoionization, cosmic-ray ionization, radiative recombination, and helium dielectronic recombination.  We use the electron-impact ionization coefficients of \citet{Voronov97} and recombination fits from \citet{VernerFerland96} and \citet{HuiGnedin97}. ${\rm H}_2$ forms via the three-body reaction, on dust grains, and via the ${\rm H}^-$ channel. It is destroyed collisionally, via reactions with ions, via LW dissociation, and via cosmic rays. ${\rm H}^-$ is evaluated via a balance of formation and destruction rates. The molecular cooling and reaction prescriptions come from \citet{GloverJappsen07} and \citet{GloverAbel08}, with updated associative-detachment and three-body formation rates from \citet{Kreckel+10} and \citet{Forrey13}.

Electrons are not explicitly transported, but are computed from (nonequilibrium) hydrogen and helium ions, and a reduced charge closure from metals and unresolved molecules. Metal abundances in this model are prescribed, rather than evolved, in the present version of DISPATCH.

We use a composition-dependent equation of state with molecular
rotational and vibrational degrees of freedom \citep{Boley+07, Vaidya+15}, similar to the molecular thermodynamic treatment in STARFORGE \citep{Grudic+20}. The pressure is the ideal translational pressure of the mixture,
\begin{align}
 P&=\frac{\rho k_{\rm B}T}{\mu m_{\rm u}},\\
 \mu^{-1}&=X\left(1-\frac{f_2}{2}+x_{\rm e}\right)+\frac{Y}{4},
 \label{eq:pressure}
\end{align}
where $m_{\rm u}$ is the atomic mass unit. The thermal specific
internal energy additionally includes the rotational and vibrational
energy of H$_2$,
\begin{equation}
 u(T,\boldsymbol{x})=
 \frac{3k_{\rm B}T}{2\mu m_{\rm u}}
 +\frac{Xf_2}{2m_{\rm u}}
 \left[\epsilon_{\rm rot}(T)+\epsilon_{\rm vib}(T)\right].
 \label{eq:thermal-eos}
\end{equation}
Here $\epsilon_{\rm rot}$ and $\epsilon_{\rm vib}$ are the mean
excitation energies per molecule. We define $n_{\rm H}=X\rho/m_{\rm u}$,
$f_2=2n_{{\rm H}_2}/n_{\rm H}$, and $x_{\rm e}=n_{\rm e}/n_{\rm H}$. The molecular partition functions use a fixed ortho-to-para ratio of 3:1. The molecular energy is tabulated from the partition functions,
and its heat capacity is obtained from the derivative of the same
interpolant. At fixed composition,
\begin{equation}
 c_v=\left(\frac{\partial u}{\partial T}\right)_{\boldsymbol{x}},
 \qquad
 \Gamma_1=1+\frac{k_{\rm B}}{\mu m_{\rm u}c_v},
 \qquad
 c_s^2=\Gamma_1\frac{P}{\rho}.
 \label{eq:gamma1}
\end{equation}
Temperature is recovered by inverting equation~\ref{eq:thermal-eos}. In our nonequilibrium network, ionization energy and molecular binding energy enter through the reaction energy terms. Neither contribution is included in the thermal energy stored by the EOS.

\subsection{Gas Heating and Cooling}
\label{sec:heating-cooling}

We evolve nonequilibrium hydrogen and helium chemistry while using reduced prescriptions for metal cooling. Related combinations of chemical networks and reduced cooling treatments are used in other multiphase models \citep{Smith+17,Hopkins+23,Kim+23}. Compression and shock heating come from the MHD update. The local term includes atomic excitation and ionization, recombination, free-free emission, molecular line cooling, photoelectric heating, reaction heating and cooling, cosmic rays, compton heating from the CMB, and dust-gas coupling. H$_2$ formation and destruction heating and cooling are tracked, and exchange energy with the gas. This tracks formation and destruction rates separately, and thereby tracks the energy of molecular cycling. The metal contribution in cold gas includes C$^+$ fine-structure cooling, ground transitions of neutral oxygen, and an aggregate molecular-metal cooling term.

Atomic cooling includes collisional excitation and ionization of
hydrogen and helium, radiative and dielectronic recombination, and
free--free emission. Coefficients for excitation come from ground-state data in CHIANTI \citep{Dufresne+24}. Collisional ionization losses use \citet{Voronov97}, recombination cooling follows \citet{HuiGnedin97} and \citet{Rosdahl+13}, and free--free emission uses \citet{Katz+96}. Photoionization heats the gas via the absorbed energy in excess of the ionizing threshold. 

At high temperatures, metals use a metal-only cooling table derived from the low-density, optically thin CIE treatment of \citet{Stofanova+21}. Hydrogen and helium losses are evaluated separately, and the metal contribution is scaled according to the electron abundance and the adopted metal abundance. This high-temperature cooling is smoothly joined to the low-temperature prescription between $20,000$ and $35,000$~K.  

\subsection{Dust and Gas--Dust Energy Exchange}
\label{sec:dust}

Dust couples radiation to the gas via absorption, re-emission, and collisions. Dust opacities in our simulation are split into two models: one for the direct emission from stars, and one for the diffuse emission, with separate spectral models appropriate to their respective applications. These are unified by a common dust abundance and temperature. Dust thermal inertia is neglected, and the dust temperature is determined by
\begin{align}
 \mathcal G &= A_{\rm gd}(T-T_{\rm d}),\\
 \mathcal G+\mathcal H_{\rm d}
 &=c\left(\alpha_{\rm em}a_{\rm R}T_{\rm d}^4
          -\alpha_{\rm abs}E_{\rm IR}\right),
 \label{eq:dust-balance}
\end{align}
where $A_{\rm gd}$ is the gas--dust collisional coupling coefficient. The coefficients $\alpha_{\rm em}$ and $\alpha_{\rm abs}$ describe emission and absorption per unit length. For M1, they are Planck means weighted at $T_{\rm d}$ and $T_{\rm col}$, respectively. For stationary angular transport, they use the blended means defined in equation~\ref{eq:ray-opacity-blend} at those same weighting temperatures. Both use the local grain composition and surviving
dust abundance. Within each computational cell, the gas and infrared radiation energies are updated together, while the dust temperature is determined from the balance between collisional energy exchange, radiative absorption, and emission.

For point-source stellar radiation, we use the dust prescription in \citet{WeingartnerDraine01}, with the revised optical properties and normalization of \citet{Draine03}, using their $R_V=5.5$ carbonaceous--silicate grain model. The absorption cross-section comes from their tabulated extinction cross-section and single-scattering albedo. The source-averaged value of this cross-section for a given band is:

\begin{equation}
  \overline{\sigma}_{{\rm d},b}(T_\star)
  =
  \frac{\displaystyle\int_b
    \sigma_{{\rm d},{\rm abs}}(\nu)W_\nu(T_\star)\,d\nu}
  {\displaystyle\int_b W_\nu(T_\star)\,d\nu}.
  \label{eq:dust-band-opacity}
\end{equation}

Ionizing bands have a blackbody source spectrum with photon-number weighting, and nonionizing bands have a spectrum using energy weighting, allowing sources with different temperatures to exhibit different spectra, and therefore dust absorption coefficients. 

For diffuse thermal radiation, we use Semenov opacities \citep{Semenov+03} taking their frequency-dependent composite-aggregate model with normal silicate composition. M1 uses Planck means for absorption and emission and a Rosseland transport mean for flux damping. Stationary angular transport instead uses the Planck--Rosseland interpolation of equation~\ref{eq:ray-opacity-blend}, consistently in attenuation, emission, and momentum coupling. Absorption and transport are weighted at $T_{\rm col}$, while emission is weighted at $T_{\rm d}$. These spectral weighting temperatures are distinct from the local dust temperatures that determine grain phase and survival. Grain compositional changes are accounted for, including the density dependence of ice evaporation, and we smooth transitions in between grain compositions. The dust contribution is reduced gradually to $0$ near the sublimation temperature of $1500$~K, to prevent a sharp drop-off at that temperature. Opacities are scaled to surviving grain mass.

\subsection{Stellar Accretion, Evolution, and Radiation Sources}
\label{sec:stellar-sources}

Sink particles represent unresolved collapsing regions, along with their host star. Formation requires a collapsing region which satisfies the Jeans criterion on the finest simulation refinement level. Refinement is triggered when the local Jeans length is resolved by fewer than 14 cells, while sink creation is triggered when the Jeans length falls below 7 cell widths at the maximum refinement level.
The candidate is the densest cell in that patch. Creation requires $\boldsymbol{\nabla}\cdot\boldsymbol{v}<0$, an exclusion radius of $10$ cells from any existing sink, and a minimum in the gravitational potential smoothed over $3$ cells, in a stencil constructed from the nearest $5$ cells.

Accretion is performed at the following rate, from cells within 5 cells of the sink:

\begin{equation}
\omega_{\rm acc}
= 0.01\sqrt{\frac{GM_\star}{r_{\rm acc}^{3}}}
+ \frac{0.1}{t_{\rm ff}}
  \max\!\left(0,1-\frac{M_\star}{M_{\rm g}}\right).
\label{eq:sink-accretion-rate}
\end{equation}
Here $t_{\rm ff}=(3\pi/32G\rho_{\max})^{1/2}$.

For the non-jet IMF runs, the fraction of gas selected from a cell at radius $r<r_{\rm acc}$ in a sink timestep $\Delta t$ is
\begin{equation}
\begin{split}
f_{\rm rem}=\max\Biggl[
&\frac{\omega_{\rm acc}\Delta t}
 {1+\omega_{\rm acc}\Delta t/0.5}
 \left(1-\frac{r^2}{r_{\rm acc}^2}\right)^2
\\
&{}\times\left(1-e^{-\rho/\bar\rho}\right),
\;1-\frac{\rho_{\rm sink}}{\rho}
\Biggr].
\end{split}
\label{eq:sink-removal}
\end{equation}
and is zero outside this radius. 
The density-dependent factor suppresses accretion from low-density gas, while the radial taper reduces mass removal near the accretion boundary.

Mass and linear momentum are transferred between the sink and the accreted gas. In runs without protostellar outflows, $50\%$ of the accreted gas is discarded along with its momentum, to represent unresolved distant ejecta. In runs with outflows, this is not included. 

Each star carries an effective temperature and radius specified by the stellar-evolution prescription used. The implementation supports a one-zone protostellar model similar to those developed by \citet{Offner+09}, or tabulated accreting protostellar models from e.g. MESA accompanied by a subsequent stellar evolution table. These prescriptions are properties of the simulation at hand. The same protostellar radius is used consistently for all feedback calculations. The stellar spectrum is calculated using a blackbody based on the effective temperature, and also includes an infrared term based on the accretion luminosity.

\subsection{Protostellar Jets}
\label{sec:jets}

In standard runs, with resolution that does not resolve the jet launching scale, jets must be treated with a subgrid model, described here. Following \citet{Cunningham+11}, we prescribe the launch process, using the simple parameterization reproduced in what follows.  Given an accreted mass $\Delta M_{\rm in}$, an outflow fraction $f_{\rm w}=0.3$ and a jet velocity factor $f_v=0.3$,
\begin{equation}
 \Delta M_\star=(1-f_{\rm w})\Delta M_{\rm in},\qquad
 \Delta M_{\rm w}=f_{\rm w}\Delta M_{\rm in},
 \label{eq:jet-mass}
\end{equation}
where $\Delta M_\star$ is the amount of accreted gas that is used for sink growth. The launch speed is
\begin{equation}
 v_{\rm w}=f_v\left(\frac{GM_\star}{R_\star}\right)^{1/2}.
 \label{eq:jet-speed}
\end{equation}
$R_\star$ is given by the stellar evolutionary model chosen for the run.

The outflow axis is chosen based on the angular momentum of the accreted gas, and, following the reference implementation, we deposit the ejecta once per local patch timestep within a shell at six to ten finest cell-widths from the sink. The deposition spatial kernel within that shell is 

\begin{equation}
 W(r,\theta)\propto
 \frac{1}{r^2\left(\sin^2\theta+\theta_0^2\right)},
 \label{eq:jet-kernel}
\end{equation}
where $\theta$ is measured from the outflow axis and $\theta_0=0.01$ sets the angular core of the launch distribution. Note that $\theta_0$ is a very flexible parameter, as at the angular cell resolution near the sink it is effectively unresolved. The jet gas is prescribed to have a temperature of $10^4$~K, and does not contribute to the magnetic field strength.

\section{Validation}
\label{sec:validation}

\subsection{Thermochemical Evolution and Equilibria}
\label{sec:thermochemistry-tests}

We test the thermochemical evolution in our simulations at fixed gas density, following the equilibrium and cooling tests of \citet{Smith+17}. The tests separately examine atomic equilibrium, cooling and recombination, molecular formation, and photodissociation. They use the simulation's reaction network and composition-dependent equation of state, including the energy exchange associated with molecular reactions, without MHD evolution or radiation transport.

\begin{figure*}[t]
\centering
\includegraphics[width=\textwidth]
{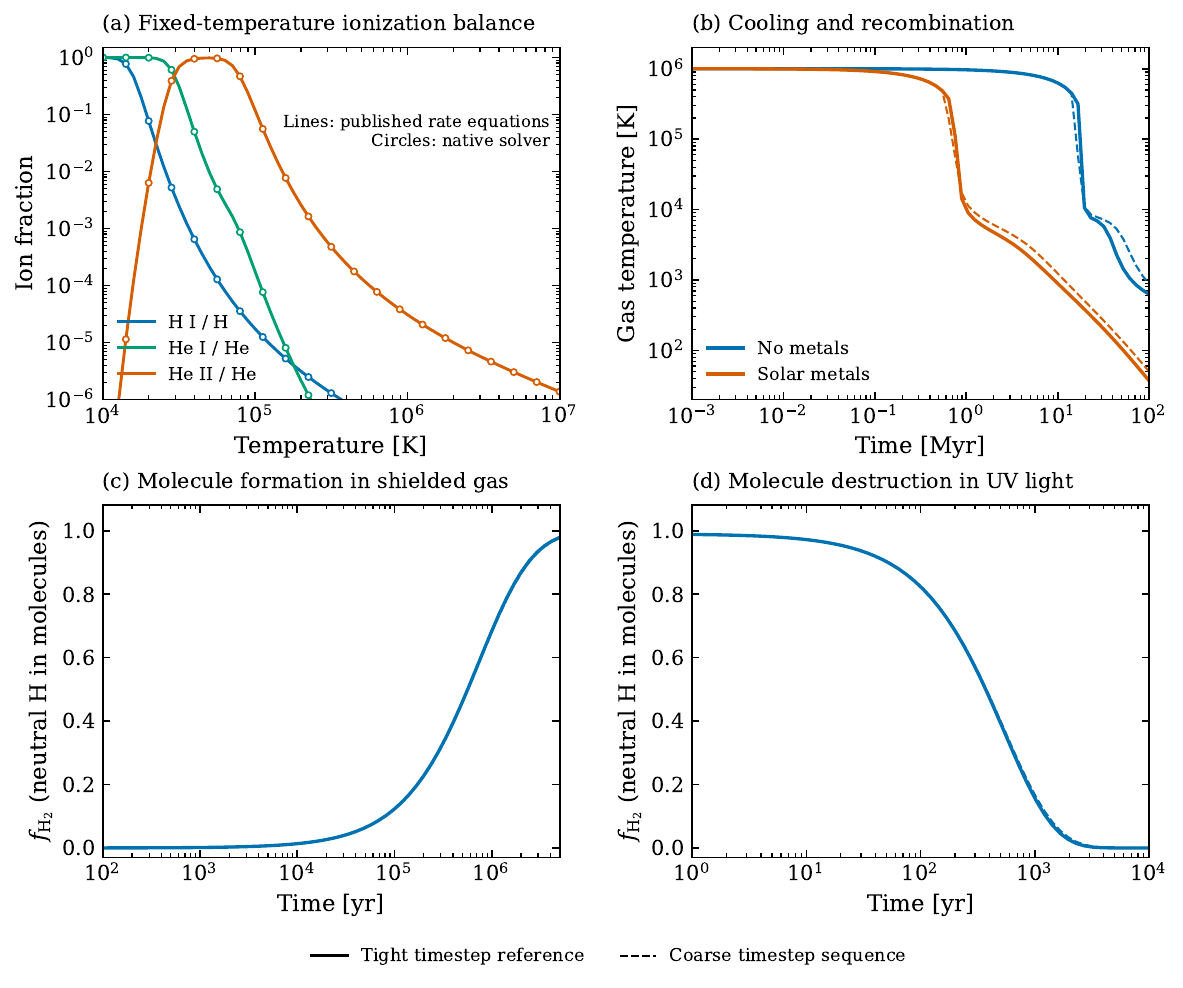}
\caption{
Thermochemical equilibrium and evolution tests. (a) Fixed-temperature H/He equilibrium fractions: independently evaluated rate equations (lines) and numerical solutions (circles), normalized to each element. (b) Constant-density cooling from $10^6\,{\rm K}$ without metals (blue) and with solar metal abundance (orange). (c) Molecular formation in initially atomic gas with solar dust and no external radiation. (d) Molecular destruction in an unshielded ultraviolet
field with $G_0=1.7$. The molecular fractions in (c) and (d) are normalized to the neutral hydrogen reservoir. The coarse run (dashed in (b)-(d)) takes one step per logarithmic output interval, with spacing $\Delta\log_{10}t=0.071$ in (b) and (d), and $0.078$ in (c); the fine timestep test divides each interval into 1024 equal-duration steps. These curves differ only in time resolution; temperature and chemical abundances evolve together.
}
\label{fig:thermochemical-validation}
\end{figure*}

At fixed temperatures between $10^4$ and $10^7\,{\rm K}$, the hydrogen and helium abundances are compared with collisional ionization equilibria obtained by independently evaluating the adopted rate coefficients. Figure~\ref{fig:thermochemical-validation}(a) shows agreement across the ionization transitions.

For coupled cooling and recombination, in panel (b) we initialize nearly fully ionized hydrogen and doubly ionized helium at $T=10^6\,{\rm K}$ and $\rho=0.1\,m_{\rm u}\,{\rm cm}^{-3}$, and evolve the gas for $100\,{\rm Myr}$. We compare zero and solar metal abundances without dust, external radiation, or cosmic rays. Both tests follow nonequilibrium H/He abundances. We test sensitivity to the time resolution, with dashed lines showing a coarse run that takes one step per logarithmic output interval, with spacing $\Delta\log_{10}t=0.071$, and solid lines showing a fine run with steps spaced at $1024\times$ higher resolution. The latter only significantly diverges from the coarse run in some later parts of the no-metal test, when the coarse timesteps are multiple Myr per step; the cooling is well-resolved in time in the typical timesteps in our simulations on the order of one to a thousand years.

We also follow molecular formation in initially atomic gas at $n_{\rm H}=10^3\,{\rm cm}^{-3}$ and $T=100\,{\rm K}$, with solar metals and dust and no external radiation or cosmic rays. Formation on grains converts the atomic reservoir into molecular hydrogen over the ensuing $5\,{\rm Myr}$ (Figure~\ref{fig:thermochemical-validation}c). Conversely, initially molecular gas at $n_{\rm H}=100\,{\rm cm}^{-3}$, $T=1000\,{\rm K}$, and molecular fraction $0.99$ is exposed to an unshielded field with $G_0=1.7$. Photodissociation removes most of the molecules within $10^4\,{\rm yr}$ (Figure~\ref{fig:thermochemical-validation}d). Temperature evolves together with the chemistry in both calculations. These plots also show coarse and fine timestepping; they agree to within $1\%$ in all cases. 

Dividing each logarithmic output interval for these histories into 256 rather than 1024 steps changes the temperature by at most $0.90\%$, $0.27\%$, $0.042\%$, and $0.014\%$ for metal-free cooling, solar-metallicity cooling, molecular formation, and molecular destruction, respectively: at time resolutions slightly coarser than our typical simulation timestepping, the thermochemistry is well-converged in time.

Separate fixed-temperature molecular tests isolate the reaction kinetics. With constant coefficients, initially atomic gas follows $f=1-\exp(-kt)$ under formation, while initially molecular gas follows $f=\exp(-kt)$ under destruction, where $f$ is the fraction of neutral hydrogen nuclei bound in molecules. Over five reaction times, the maximum absolute fraction error decreases from $1.12\times10^{-2}$ to $7.17\times10^{-4}$ as the number of timesteps increases from 40 to 640. The finite-step solutions approach equilibrium more slowly than the analytic exponentials, as expected for the backward-Euler molecular update.

\subsection{Hot Gas and Metal Cooling}
\label{sec:hot-cooling-tests}

We compare our optically thin, low-density cooling coefficients with the independent collisional ionization equilibrium (CIE) prescription of \citet{GnatFerland12}. Their element-resolved coefficients are combined using the same helium and metal abundances as our model. Both cooling functions are expressed per $n_{\rm e}n_{\rm H}$, with dust, external radiation, and cosmic rays excluded. Hydrogen and helium are placed in CIE for this comparison, rather than evolved out of equilibrium as in the preceding tests.

\begin{figure*}[t]
\centering
\includegraphics[width=\textwidth]{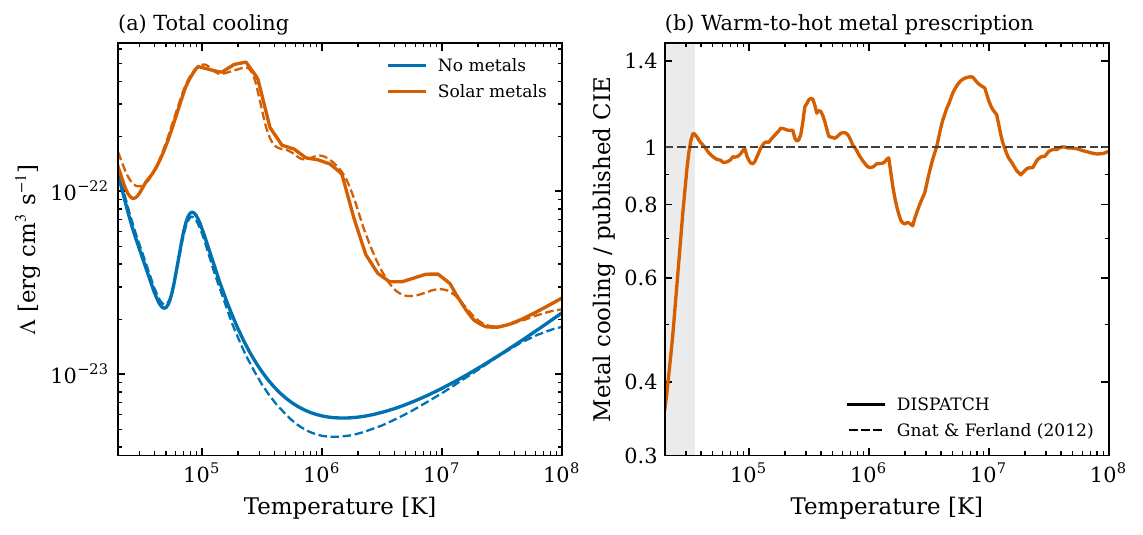}
\caption{
Low-density cooling at matched elemental abundances. Left: total cooling coefficients without metals (blue) and at solar metal abundance (orange). Solid curves show our prescription, and dashed curves show the CIE calculations of \citet{GnatFerland12}, with both expressed per $n_{\rm e}n_{\rm H}$. Right: our metal cooling coefficient divided by the published reference. The shaded interval marks the transition region between the low-temperature prescription and the hot-gas table. 
}
\label{fig:hot-gas-cooling}
\end{figure*}

Figure~\ref{fig:hot-gas-cooling} shows the total cooling with zero and solar metal abundance, together with the ratio of our metal cooling to the published reference. Above $3.5\times10^4\,{\rm K}$ and below $10^8\,{\rm K}$, where our hot-gas prescription is fully active, this ratio ranges from 0.74 to 1.3. Iron is the main source of the largest differences, near $2.3\times10^6$ and $7.1\times10^6\,{\rm K}$, which are prescription-based and come from differences between the SPEX-based prescription adopted here \citep{Stofanova+21} and the independent CLOUDY-based reference. At lower temperatures $<2\times10^5\, {\rm K}$ more typical outside of supernovas, but above the transition temperature, the models agree well.

The transition between the low-temperature metal prescription and the hot table is continuous over $2\times10^4$--$3.5\times10^4\,{\rm K}$. At $2\times10^4\,{\rm K}$, metal cooling is $0.36$ times the CIE reference, but total cooling, including hydrogen and helium, is only $17\%$ lower. The deficit principally reflects omitted Fe, Mg, and Si channels in the reduced nebular treatment adopted following \citet{Kim+23}. These elements are strongly depleted in dust-bearing H\,II regions \citep{PeimbertPeimbert10}, and adding their CIE cooling would not generally improve a photoionized-gas prescription. For dense shocked gas, the measured difference increases the instantaneous cooling time by approximately $20\%$ while leaving it short compared with the resolved flow time. We therefore expect a modest effect on bulk cloud dynamics, with greater sensitivity in warm cooling-layer structure and emission, particularly where grain destruction returns refractory elements to the gas.

\subsection{Diffuse Radiation Transport and Coupling}
\label{sec:diffuse-tests}

We test diffuse radiation using the temperature structure of an irradiated dust envelope, the relaxation of a radiation perturbation, and the resulting momentum transfer to gas. This allows us to test the spectral approximation, the accuracy of the radiation transport and coupling, and the radiation pressure accuracy.

We first consider a spherical envelope with $\rho\propto r^{-2}$, inner and outer radii of $100$ and $1000\,{\rm AU}$, and a central $6000\,{\rm K}$ blackbody source. Following the spherical-envelope benchmark approach of \citet{Ivezic+97}, we compare with a frequency-dependent solution obtained using DUSTY \citep{IvezicElitzur97}. The radial absorption optical depths at $1\,\mu{\rm m}$ are $\tau_1=$ $0.1$, $1$, and $10$. For each optical depth, the source luminosity is chosen to give an inner dust temperature of $100\,{\rm K}$ in the reference solution. Both tests use the $R_V=5.5$ Draine absorption spectrum for stellar and thermal radiation, with scattering omitted, which matches the prescription of our comparison. Hydrodynamics, chemistry, gas--dust collisional exchange, and dust destruction are disabled. The envelope is embedded in a $3200\,{\rm AU}$ box with spatially varied AMR and a finest cell width of $6.25\,{\rm AU}$. We compare both diffuse solvers: M1 with $\widehat c=30\,{\rm km\,s^{-1}}$, and stationary transport using 146 angular directions.

\begin{figure*}[t]
\centering
\includegraphics[width=\textwidth]{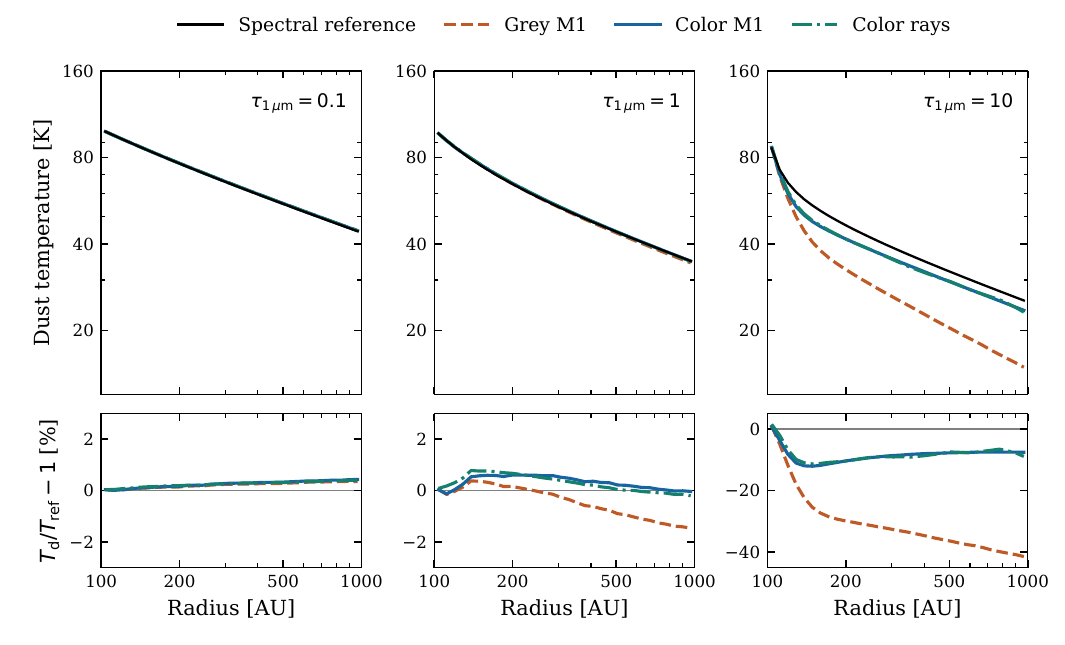}
\caption{
Dust temperatures in irradiated spherical envelopes with $\rho\propto r^{-2}$. Columns show radial absorption optical depths at $1\,\mu{\rm m}$ of $0.1$, $1$, and $10$. Black curves give the frequency-dependent DUSTY reference. Blue and green curves show the color-dependent M1 and diffuse ray results. Orange dashed curves show, for comparison, an alternate M1 implementation that uses the local dust temperature for its color, labelled grey M1. Profiles are volume-weighted radial medians, and the lower panels show their fractional differences from the reference evaluated at the same radii.
}
\label{fig:diffuse-envelope}
\end{figure*}

Figure~\ref{fig:diffuse-envelope} shows that both solvers reproduce the reference radial temperature profile to within $1\%$ for $\tau_1\leq1$. At $\tau_1=10$, the maximum radial-median temperature differences are $12.1\%$ for M1 and $18.5\%$ for stationary rays. The difference in the high-optical-depth case has an identifiable physical origin: our simulated temperature spectrum. As shown in our comparison between our color M1 and grey M1 models, our color temperature significantly improves dust temperature accuracy at high optical depths. However, a single color temperature does not fully describe the spectrum of reprocessed dust emission, which is weighted by the frequency-dependent emissivity. In the outer part of the $\tau_1 =10$ M1 envelope, the radiation energy density is within $3\%$ of the spectral reference, but the effective absorption opacity is about $37\%$ lower. The resulting underheating accounts for the significant temperature deficit there.

We also test M1 transport and material exchange using a small ($1\%$) sinusoidal perturbation in radiation energy,
\begin{equation}
\begin{aligned}
E_{\rm IR}(x,0)
&= E_0\left[1+0.01\cos\left(\frac{2\pi x}{L}\right)\right],\\
E_0 &= a_{\rm R}(100\,{\rm K})^4.
\end{aligned}
\end{equation}
with initially vanishing radiation flux and gas and dust temperatures of $100\,{\rm K}$. The periodic domain has $L=1.0001\times10^{17}\,{\rm cm}$ and $\rho=10^{-15}\,{\rm g\,cm^{-3}}$, which corresponds to cell optical depths of $6.8$ and $13.6$ on its two AMR levels. Density and composition are held fixed. An independent solution is obtained by integrating the linearized moment and gas--dust exchange equations using the background opacity, heat capacity, and collisional coupling coefficient, with the same reduced light speed as the simulation.

\begin{figure*}[t]
\centering
\includegraphics[width=\textwidth]{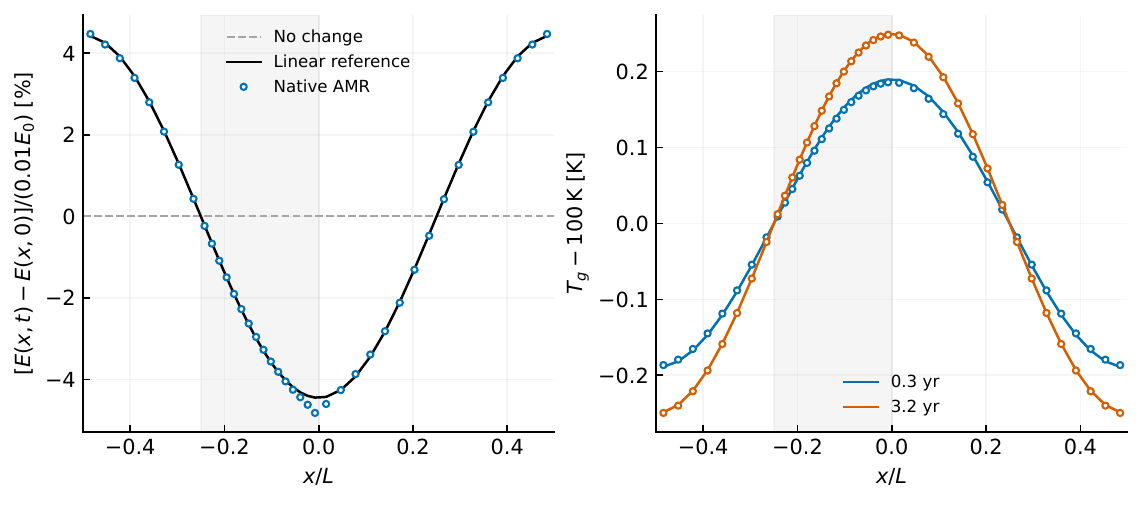}
\caption{
Diffuse radiation transport and gas--dust relaxation on a spatially varied AMR mesh. Left: change in radiation energy from the initial profile at $1585\,{\rm yr}$, normalized by the imposed perturbation amplitude $0.01E_0$. The dashed line denotes no evolution. Right: gas-temperature perturbations at $0.32$ and $3.2\,{\rm yr}$, obtained with a maximum patch timestep of $0.016\,{\rm yr}$. Symbols show the numerical solution and solid lines show the analytic solution, including the adopted reduced light speed. Shading identifies the finer mesh.
}
\label{fig:diffuse-transport}
\end{figure*}

Figure~\ref{fig:diffuse-transport} shows the redistribution of radiation and the initial material relaxation. At $1585\,{\rm yr}$, the volume-weighted mean absolute radiation-energy error is $0.04\%$, and the mean gas-temperature error is of order $10^{-4} \%$. During the initial relaxation, halving the maximum patch timestep from $0.032$ to $0.016\,{\rm yr}$ reduces the mean temperature error at
$0.32\,{\rm yr}$ by a factor of two, consistent with the first-order temporal update. A corresponding optically thin test, with cell optical depths of $0.07$--$0.14$, has a radiation-energy error of $0.40\%$ of the initial perturbation amplitude at $15.85\,{\rm yr}$.

\begin{figure}[t]
\centering
\includegraphics[width=\columnwidth]{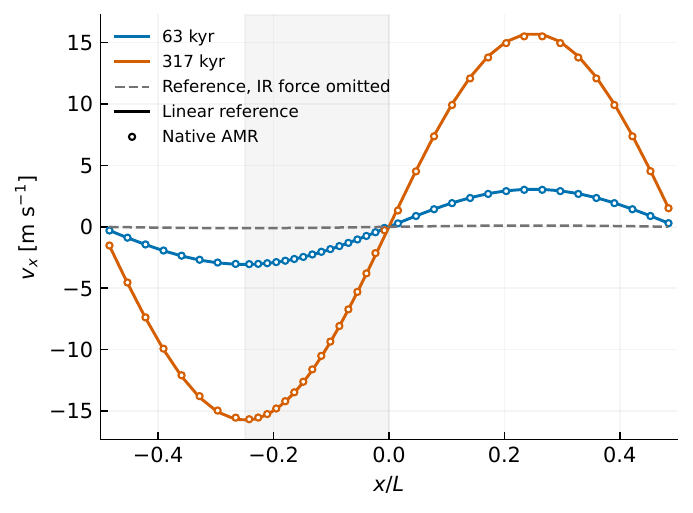}
\caption{
Gas acceleration by diffuse radiation at $63$ and $317\,{\rm kyr}$. Circles show the numerical velocity and solid lines the linearized solution including gas pressure and radiation force. The dashed curve is the late-time reference with radiation force omitted. Shading identifies the finer mesh.
}
\label{fig:diffuse-force}
\end{figure}

Finally, we test diffuse radiation pressure by allowing the gas to respond dynamically to the same radiation perturbation. We use $\rho=10^{-18}\,{\rm g\,cm^{-3}}$ and $L=10^{20}\,{\rm cm}$, preserving the cell optical depths. The reference includes continuity, gas pressure, material exchange, and the radiation force $\alpha_R F_x/c$. At $317\,{\rm kyr}$, the mean velocity error is less than $1\%$ (Figure~\ref{fig:diffuse-force}).

\subsection{Radiative Collapse of a Prestellar Core}
\label{sec:collapse-test}

To test the transition from nearly isothermal collapse to adiabatic heating, we simulate a uniform, gravitationally unstable core with mass $1\,M_\odot$ and initial density $5.3\times10^{-18}\,\mathrm{g\,cm^{-3}}$, following the setup used by \citet{Grudic+20}. The gas and radiation background are initialized at $10\,\mathrm{K}$, with solar metallicity and an initial dust-to-gas mass ratio of $0.01$. We evolve self-gravity and coupled gas--dust--radiation thermodynamics, using our fiducial settings. Radiation pressure is disabled.

\begin{figure}[t]
\centering
\includegraphics[width=\columnwidth]{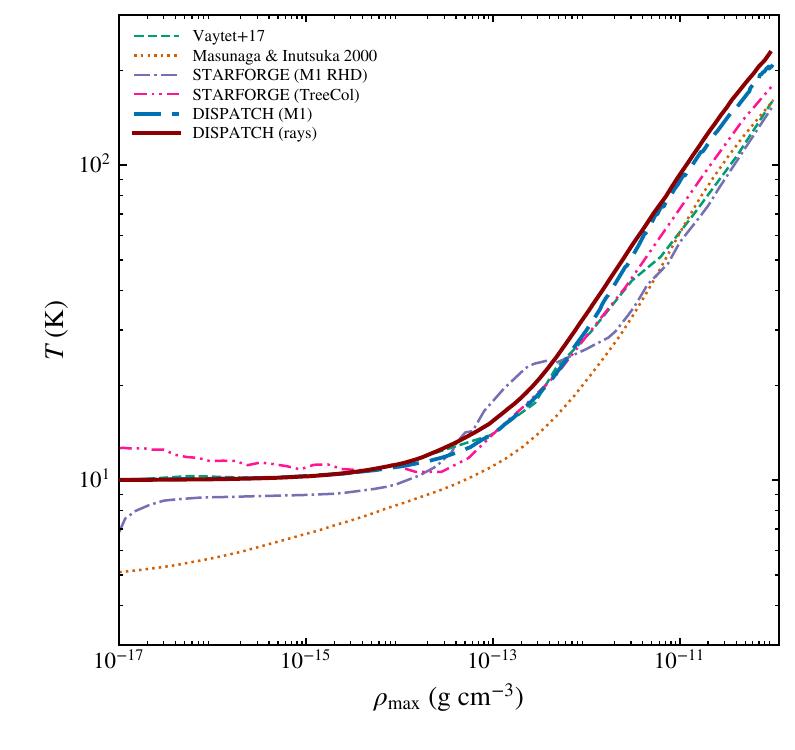}
\caption{
Temperature of the densest gas cell versus peak density during collapse of an initially $10\,\mathrm{K}$, $1\,M_\odot$ core. The blue dashed curve shows our simulation with M1 diffuse radiation; the brown curve shows it with diffuse rays. Other curves show the STARFORGE M1 and TreeCol results, together with the \citet{Masunaga+00}, \citet{Vaytet+17}, and \citet{Grudic+20} comparisons.
}
\label{fig:collapse-test}
\end{figure}

Figure~\ref{fig:collapse-test} shows the temperature of the densest gas cell as a function of peak density, for both M1 diffuse radiation (blue dashed) and diffuse rays (brown line). The simulation recovers the initially nearly isothermal evolution, followed by heating around $10^{-13}\,\mathrm{g\,cm^{-3}}$ as radiative cooling becomes insufficient to remove the compressional energy. The hotter post-isothermal trajectory compared to references reflects differences in the molecular thermodynamics. We adopt a frozen $3:1$ ortho-to-para H$_2$ mixture, representing the statistical formation ratio retained when spin conversion is slow \footnote{Recent STARFORGE simulations use an equation of state similar to that adopted here with the same assumption, whereas their simulations shown in this figure used a somewhat different treatment.}. This avoids assuming instantaneous spin equilibrium, but neglects conversion during the preceding cloud evolution. The assumption has little influence on the initial cold thermal response, while the subsequent heating trajectory depends appreciably on the spin mixture \citep{Vaytet+14}. For the adopted mixture, we calculate internal energy and heat capacity consistently from molecular partition functions, rather than using a fitted adiabatic index to convert internal energy to temperature \citep{Boley+07}. In a matched collapse control, substituting the historical public GIZMO energy--temperature prescription lowered the temperature at $\rho_{\max}=10^{-12}\,{\rm g\,cm^{-3}}$ from $31$ to $23\,{\rm K}$ and recovered a cooler plateau. This treatment therefore provides a thermodynamically consistent evolution for the adopted mixture, and the control demonstrates its substantial effect on the collapse trajectory. The assumed spin ratio remains a physical approximation, particularly for gas whose ratio has evolved before collapse.

\subsection{Radiation Pressure-driven Expansion}
\label{sec:radiation-pressure-test}

To test our direct radiation pressure prescription, we place a central source of luminosity $L_\star=10^6\,L_\odot$ in initially stationary, uniform gas with density
$\rho_0=10^{-21}\,\mathrm{g\,cm^{-3}}$. The gas is contained in a $32\,\mathrm{pc}$ cube with a fixed, spatially varying AMR hierarchy with cell widths between $1$ and $0.25\,\mathrm{pc}$. The gas is isothermal with sound speed $0.2\,\mathrm{km\,s^{-1}}$. Radiation is transported in a single purely absorbing band; gravity, photoheating, and diffuse radiation are disabled. This isolates the momentum coupling tested in \citet{Grudic+20}. We choose constant opacities giving initial box optical depths $\tau_{\rm box}=\kappa\rho_0L_{\rm box} =19.2$ and $192$, corresponding to optical depths of $0.15$ and $1.5$ across the finest cells.

\begin{figure}[t]
\centering
\includegraphics[width=\columnwidth]{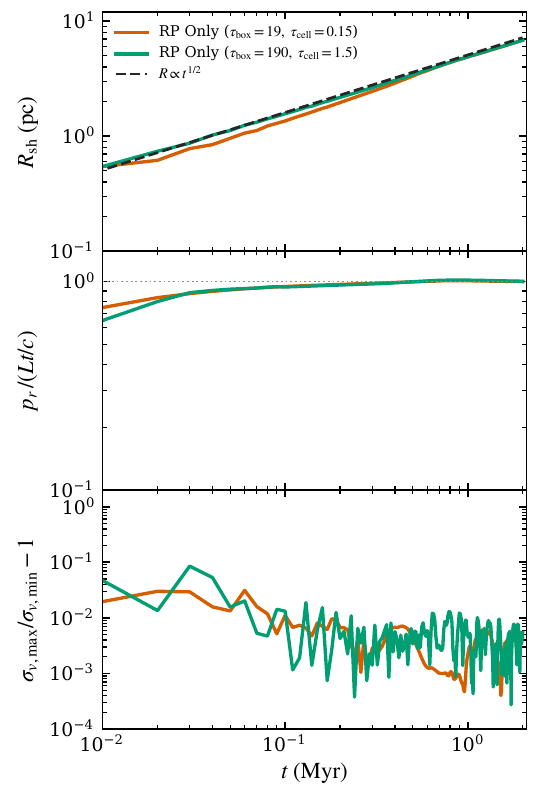}
\caption{
Expansion driven solely by direct radiation pressure for two absorption opacities. Legend values of $\tau_{\rm cell}$ refer to the finest initial cells. \textit{Top:} shell radius, defined as the median radius of the cells with $\rho>1.01\rho_0$. The dashed line shows equation~\ref{eq:radiation-shell}. \textit{Middle:} total radial gas momentum in units of the emitted photon momentum $L_\star t/c$; the dotted line marks unity. \textit{Bottom:} numerical velocity anisotropy.
}
\label{fig:radiation-pressure-shell}
\end{figure}

For an opaque, geometrically thin shell sweeping up the ambient gas with negligible thermal pressure, momentum conservation gives 
\begin{equation}
\frac{d}{dt}\left(\frac{4\pi}{3}\rho_0R^3\dot R\right)
=\frac{L_\star}{c},
\qquad
R(t)=\left(\frac{3L_\star t^2}{2\pi\rho_0c}\right)^{1/4}.
\label{eq:radiation-shell}
\end{equation}
Figure~\ref{fig:radiation-pressure-shell} compares the numerical evolution with this prediction. Both runs approach the expected $R\propto t^{1/2}$ expansion, and the velocity anisotropy and momentum budget agree to within 1\% with the analytic result. These results demonstrate late-time momentum coupling for both resolved and unresolved initial absorption lengths on the adopted AMR hierarchy.

\subsection{Overlapping Point-source Illumination}
\label{sec:interacting-point-sources}

To demonstrate our hybrid radiation method's power in dense streams with multiple light sources, we show a test illumination of a clump, optically thick to ionizing point-source radiation and moderately optically thin to infrared re-emission. The density in the test is held fixed, and gas motion and gas--dust collisional exchange are disabled to isolate the radiation and dust response. This aims to isolate the effects of our ray-tracing point-source mechanism and our diffuse M1 method as well as showing their coupling in separate regions.

We adopt a constant absorption opacity for stellar radiation and,
for this test, constant infrared means. The optical depths through the clump diameter are $19.9$ and $0.311$ for stellar and infrared radiation, respectively. Absorbed stellar energy is re-emitted by dust into the diffuse M1 field, evolved with $\widehat c=30\,\mathrm{km\,s^{-1}}$. The fixed AMR hierarchy has cell widths of $31.2$, $15.6$, $7.8$, and $3.9\,\mathrm{AU}$, with the finest cells in a slab containing the sources and the plotted slice. This test uses constant infrared opacities and follows dust temperature.

\begin{figure*}[t]
\centering
\includegraphics[width=\textwidth]{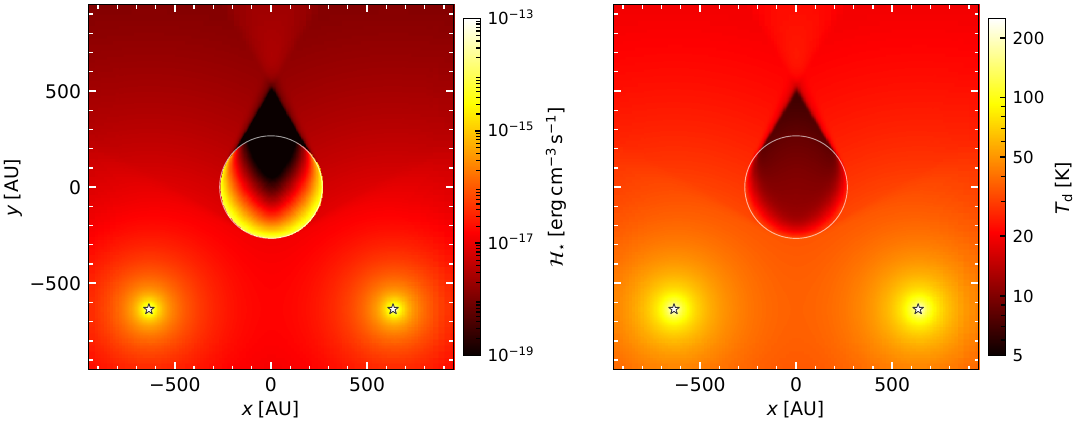}
\caption{
Two point sources irradiating a dense clump. Slices through the source plane at $t=450\,\mathrm{yr}$ show the absorbed stellar power per unit volume $\mathcal H_\star$ (left) and dust temperature $T_{\rm d}$ (right). Stars mark the sources and the circle marks the clump boundary; coordinates are relative to the clump center. Direct irradiation produces overlapping shadows, while infrared re-emission smooths the dust temperature distribution.
}
\label{fig:interacting-point-sources}
\end{figure*}

Figure~\ref{fig:interacting-point-sources} shows the absorbed stellar power and dust temperature. The dual source produce the expected umbra and penumbra. Diffuse re-emission produces a smoother temperature distribution and heats material shielded from direct stellar illumination. This demonstrates the capability of our hybrid model of directional stellar transport and diffuse reprocessing for multiple-source illumination of dense clumps.

\subsection{Ionization Fronts and H II Region Expansion}
\label{sec:hii-tests}

\begin{figure}[t]
\centering
\includegraphics[width=\columnwidth]
{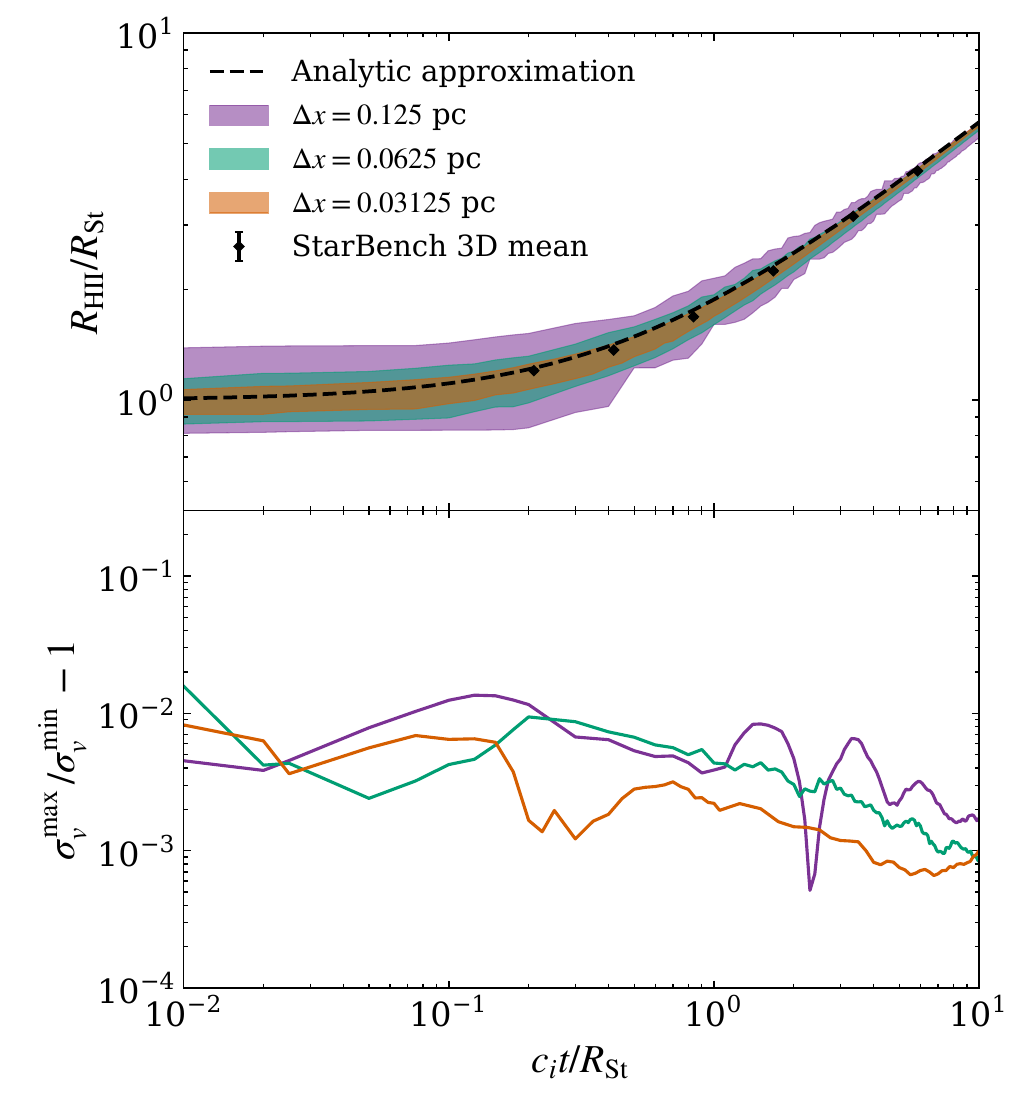}
\caption{
Pressure-driven expansion of an HII region at three spatial resolutions. Top: shaded intervals span the radii at which the spherically averaged ionized fraction falls from $0.99$ to $0.01$; their widths measure the front thickness. The dashed curve is equation~\ref{eq:hii-expansion}, and black diamonds show the published three-dimensional StarBench means and scatter \citep{Bisbas+15}. Bottom: velocity anisotropy, $\sigma_{v,\max}/\sigma_{v,\min}-1$, where the dispersions are the square roots of the eigenvalues of the mass-weighted velocity covariance tensor. This quantity vanishes for spherical expansion.
}
\label{fig:hii-expansion}
\end{figure}

We test the coupling of ionizing radiation to hydrodynamics through the pressure-driven expansion of an HII region, following the StarBench setup \citep{Bisbas+15}. An initially stationary, uniform pure-hydrogen medium with density $\rho_0=5.21\times10^{-21}\,{\rm g\,cm^{-3}}$ fills a $16\,{\rm pc}$ box. A single O-type star emits $\dot N=10^{49}\,{\rm s^{-1}}$ monochromatic ionizing photons at $13.6\,{\rm eV}$. We adopt a constant case-B recombination coefficient $\alpha_{\rm B}=2.7\times10^{-13}\,{\rm cm^3\,s^{-1}}$. This isolates ionization-front propagation and its hydrodynamic response from the detailed thermal model. Dust and radiation pressure are omitted.

The initial Str\"omgren radius is \begin{equation}
R_{\rm St}=
\left(\frac{3\dot N}
{4\pi\alpha_{\rm B}n_{\rm H,0}^{\,2}}\right)^{1/3}
=0.313\,{\rm pc}.
\end{equation}
We compare the expansion with the Hosokawa--Inutsuka approximation, as discussed by \citet{Bisbas+15},
\begin{equation}
R_{\rm HII}(t)=R_{\rm St}
\left(1+\frac{7}{4}\sqrt{\frac{4}{3}}\,
\frac{c_{\rm i}t}{R_{\rm St}}\right)^{4/7},
\label{eq:hii-expansion}
\end{equation}
where $c_{\rm i}=12.90\,{\rm km\,s^{-1}}$ is the isothermal sound speed of the ionized gas. We test the model at $3$ simulation resolutions reflective of the range of possible resolutions encountered in our simulations, to ensure that the expansion is well-captured at any resolution. As shown, these converge well to the analytic solution, and decreasing the resolution primarily acts to spread out the ionization front. We also show in the bottom panel that the velocity anisotropy is under $1\%$ at almost all times except for some very early times at all of our resolutions. The expanding region therefore retains close spherical symmetry despite the Cartesian AMR mesh and patch-local evolution.

\subsection{Jet and Outflow Propagation}
\label{sec:jet-tests}

\begin{figure*}[t]
\centering
\includegraphics[width=0.9\textwidth]{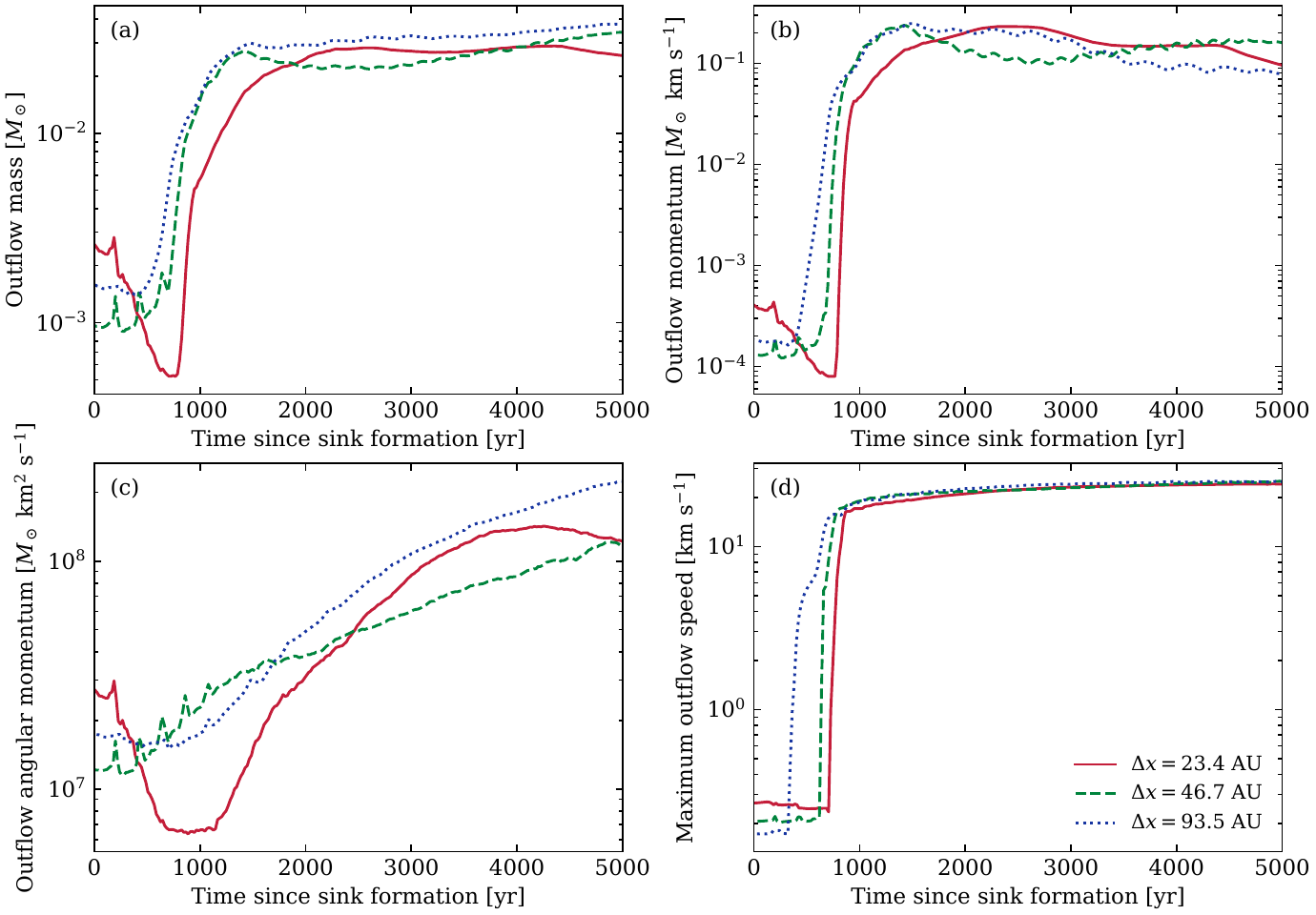}
\caption{
Resolution dependence of outflows from a collapsing, rotating, magnetized core. Panels show (a) the instantaneous outward-moving gas mass, (b) scalar momentum, (c) signed angular momentum about the launch axis, and (d) maximum sink-relative speed, measured in the two cylindrical volumes described in the text. Colors and line styles distinguish the finest cell widths. Time is measured from each run's sink formation, rather than from the onset of injection or arrival at the 1500~AU measurement window.
}
\label{fig:jet-resolution}
\end{figure*}

We examine the resolution dependence of the protostellar outflow prescription during the collapse of a rotating, magnetized core, using diagnostics similar to those of \citet{Federrath+14}. The initially uniform core has mass $1\,M_\odot$, density
$5.3\times10^{-18}\,{\rm g\,cm^{-3}}$, and temperature $10\,{\rm K}$. We impose solid-body rotation with $\Omega t_{\rm ff}=0.2$ and a uniform magnetic field of $125\,\mu{\rm G}$ parallel to the rotation axis, corresponding to a mass-to-flux ratio of $5.2$ times the critical value. This test includes self-gravity, ideal MHD, thermochemistry, and diffuse radiation via M1. Stellar irradiation and radiation pressure are disabled to isolate mechanical feedback.

We compare three AMR simulations with finest cell widths of $23.4$, $46.7$, and $93.5\,{\rm AU}$, with $46.7\,{\rm AU}$ representing our fiducial resolution. All of the tests use the outflow prescription described in Section~\ref{sec:jets}, with $f_{\rm w}=f_v=0.3$ and the launch axis fixed along the initial rotation axis. The stellar radius entering the launch speed is supplied by the one-zone protostellar model. The injection shell extends from six to ten finest cell widths, so its physical size changes with resolution. This comparison therefore tests the resolution-scaled subgrid prescription.

Following \citet{Federrath+14}, we measure the mass, scalar momentum, axial angular momentum, and maximum speed of outward-moving gas in two cylindrical volumes aligned with the outflow. Each cylinder has radius and height $1500\,{\rm AU}$, with its near base $1500\,{\rm AU}$ above or below the sink. This measurement distance lies outside the injection region at all three resolutions. Velocities and positions are measured relative to the sink, and time is measured from sink formation.

Figure~\ref{fig:jet-resolution} shows that the outflows approach similar late-time masses and speeds despite differences in their early evolution. At $2000\,{\rm yr}$ after sink formation, the fiducial and finest runs differ only mildly in outflow mass (better than $10\%$ differences) and maximum speed (circa $1\%$ differences), and differ within usable limits in linear momentum (circa $30\%$ differences). Their axial angular momenta differ significantly at times, indicating greater sensitivity of this quantity to resolution. Across all three runs, the maximum measured outflow speed is about $25\,{\rm km\,s^{-1}}$.

\subsection{Spatial and Temporal Accuracy}
 \label{sec:numerical-accuracy}
 
\begin{figure*}[t]
\centering
\includegraphics[width=\textwidth]{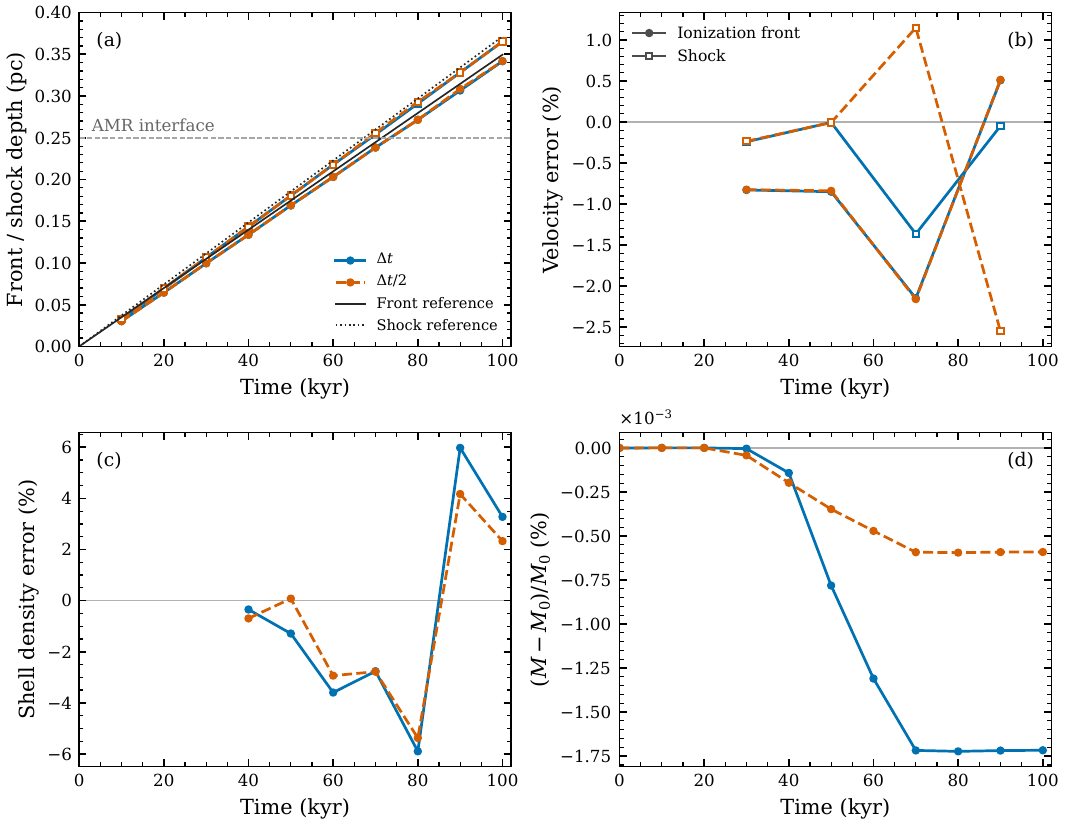}
\caption{A planar ionization front crossing a fine-to-coarse AMR interface at $0.25\,\mathrm{pc}$. Panels compare front and shock positions, speed errors, compressed-shell density, and the fractional gas-mass change, $[M_{\rm g}(t)-M_{\rm g}(0)]/M_{\rm g}(0)$, expressed as a percentage. Here $M_{\rm g}$ is the total gas mass summed over leaf-cell volumes, excluding covered coarse cells. Original and half-timestep runs are shown, demonstrating the time stability of the solution. Speed errors use common $20\,\mathrm{kyr}$ intervals; the mass panel retains all available outputs. Physical boundaries prevent gas-mass escape.}
\label{fig:async-amr}
\end{figure*}

To examine radiation propagation across asynchronous AMR boundaries, we follow a controlled planar D-type front through a fixed fine-to-coarse interface, using the problem family of \citet{KesselDeynetBurkert2000}. Initially $n_{\rm H}=10^3\,\mathrm{cm^{-3}}$, with neutral and ionized temperatures prescribed as $100$ and $10^4\,\mathrm{K}$. The incident photon flux increases as $F(t)=Ct$, with $C=5.07\times10^{-4}\,\mathrm{cm^{-2}\,s^{-2}}$. Cell widths are approximately $201$ and $403\,\mathrm{AU}$, and the interface lies $0.25\,\mathrm{pc}$ from the illuminated boundary. Patches retain ordinary local timesteps and boundary exchange, without cross-patch radiation convergence. We show the result of this test in Figure~\ref{fig:async-amr}.

Late-time jump conditions predict front and shock speeds of $3.4$ and $3.6\,\mathrm{km\,s^{-1}}$ and a shell density of $1.59\times10^4\,\mathrm{cm^{-3}}$. In the half-timestep run, late-time interval-averaged speed errors are at most $2.2\%$ and $2.5\%$, respectively; at $100\,\mathrm{kyr}$, the median shell-density error is $2.3\%$. The position reference is the late-time solution. 

\section{Discussion}
\label{sec:discussion}

We have extended DISPATCH to couple point-source radiation and stellar feedback to the thermochemical and dynamical evolution of accreting gas. The tests in  Section~\ref{sec:validation} examine complementary aspects of this coupling: the thermal
transition during gravitational collapse, radiation-driven expansion, illumination by multiple sources, and the propagation of jet outflows. Together, these allow us to follow stellar feedback and derive a self-consistent IMF, while accurately resolving illumination and shadowing from bright, ionizing sources that erode and substantially impact the evolution of accreting flows that feed high-mass stars.

\subsection{Comparison with Existing Methods}
\label{sec:method-comparison}

The current method shares a goal with STARFORGE \citep{Grudic+20}: to follow the growth of individual stars while minimizing the number of subgrid models that impact that growth, allowing tests of how different physical assumptions impact the IMF. This goal is also complementary to TORCH \citep{Wall+20,Polak+24}, which follows much of the same physics but with a prescribed IMF. In the cluster-formation prescription in the latter, individual stars are not grown over time, and accretion cannot be followed to understand the origin of the IMF.

Separating direct stellar irradiation from diffuse reprocessed radiation follows established hybrid transport approaches \citep{Kuiper+10,Klassen+14,Rosen+17}. Our implementation combines multiband short-characteristics transport of stellar radiation with either M1 or stationary angular transport of diffuse radiation, coupling these fields to nonequilibrium thermochemistry within DISPATCH's asynchronous patch-local evolution. Other codes, such as RAMSES-RT and STARFORGE, use moment transport with an M1 closure \citep{Rosdahl+13,Grudic+20}, which cannot accurately represent crossing radiation beams. In contrast, VETTAM \citep{Menon+22} uses a variable Eddington tensor to retain information about the angular distribution of radiation. Our diffuse ray method similarly preserves angular information, allowing radiation from different directions to cross without the artificial interactions associated with M1. Even when diffuse radiation is treated with M1, however, our hybrid ray-tracing/M1 method preserves the directions of radiation from individual stellar sources, allowing overlapping stellar illumination and geometric shadows. Its limitation is that crossing streams of diffuse radiation are still treated approximately by M1.


These transport choices operate within DISPATCH's patch-local evolution \citep{Nordlund+18}. Spatial refinement gives higher resolution around stars and collapsing gas, while local timesteps allow regions with different dynamical timescales to advance at different rates. This has advantages and disadvantages as a method: compared to particle-based methods, the regular mesh in the present method works naturally and stably with a simple hierarchical refinement scheme. 
Local timestepping concentrates computational effort in the most rapidly evolving regions of the box, providing substantial efficiency gains over simulations with a global timestep and additional gains over conventional hierarchical timestepping.
However, there are some tradeoffs: patch guard zone interpolation at asynchronous times is not perfectly conservative, and while as shown here and in \citet{Nordlund+18} this is often a mild effect for simulation accuracy, with the current version of the framework conservation losses must be examined explicitly for a given problem (a version under development solves this by computing explicit flux correction terms). The patch-based structure also comes with a cost in time efficiency: compared to a method that can step a handful of cells around a sink particularly quickly, we must step a minimum of one patch worth of cells at the minimum timestep.  The downside in time efficiency however comes with upside in terms of SIMD vectorization and cache efficiency.

\subsection{Implications for Star Formation and Feedback}
\label{sec:implications}

The principal application of our method is in understanding the interplay of direct ionizing radiation from stars and larger-scale accretion flows and gas features. In the inertial-inflow model, turbulent flows can sustain massive stellar growth for long periods of time, allowing sustained accretion for Myr timescales that is limited primarily by the timescale of supernova feedback \citep{Padoan+20}. 
However, the simulations underlying the inertial-inflow model do not include radiative feedback, which may disrupt accretion flows on shorter timescales, altering the duration and variability of accretion and the fraction of the inflowing gas that reaches the star.

Simulations including radiative and mechanical feedback already find prolonged massive-star growth \citep{Grudic+22,Wainer+26}. In particular, \citet{Wainer+26} associate the emergence of massive stars from their surrounding gas with the termination of accretion, interpreting this transition as feedback overcoming the local inflow. An outstanding question is how this local transition relates to the evolution of the larger-scale gas supply: how much does radiation disrupt the feeding flows, and to what extent does a declining supply allow feedback to become effective?

Directional irradiation and mechanical outflows act together in this process. Outflows can open low-density cavities, changing radiation escape and its coupling to accreting gas \citep{RosenKrumholz20}. In a cluster, the feeding flows can also be illuminated by stars other than those they supply. Following individual sources and their directions of illumination allows us to investigate where these flows are eroded and how irradiation changes the mass reaching each star.

These methods, however, are not limited to the formation of star clusters, nor small-box IMF runs. Our simulations are in the RAMSES lineage, and can interface with larger-scale RAMSES simulations, offering a route to zoom-in simulations of GMCs within their large-scale environment. Such simulations could follow how the supply of gas from the surrounding interstellar medium interacts with feedback from the cluster, connecting cloud assembly, stellar growth, and cloud dispersal within a common framework. 

The adaptive spatial resolution and local timestepping also naturally lend themselves to finer-scale studies of individual collapsing cores and their accretion flows. Such simulations would allow the protostellar environment to be studied at resolutions that are impractical throughout an entire cloud. Extending this to metal-poor and primordial star formation, with the appropriate chemical and thermal prescriptions, would provide a way to investigate how the mechanisms regulating stellar growth change as dust shielding and metal cooling become less effective. All told, this offers the prospect of studying individual stellar feedback in environments ranging from nearby molecular clouds to the sites of the first stars.

\subsection{Limitations and Range of Applicability}
\label{sec:limitations}

The radiation model retains approximations in both angle and frequency. Our ray-traced method preserves source directions, but finite spatial resolution and short-characteristic interpolation still affect shadow boundaries and ionization-front widths. We also approximate the radiation spectrum in multiple ways, including in our blackbody approximation for stellar atmospheres and in our band-averaged spectral indices. The on-the-spot approximation also omits the nonlocal transport of ionizing recombination radiation, which can affect gas shielded from direct sources. For the diffuse field, M1 cannot preserve distinct crossing streams, and our band-averaged field imperfectly retains the spectrum of mixed or diluted radiation. The stationary-ray opacity interpolation recovers the optically thin and LTE diffusion limits, but remains an approximation at intermediate optical depth. Its local column estimate is motivated by self-gravitating gas, and is less physically motivated in irradiated cavities and interfaces where the relevant column is set by nonlocal geometry.

The M1 diffuse field option also makes approximations in time, in using a reduced speed of light. The adequacy of this approximation depends on radiation transport and material equilibration, including diffusion through optically thick gas, rather than on the ionized-gas sound speed alone. In the adopted equations, closed local exchange conserves $\rho u+E_{\rm IR}/\epsilon$, so reducing $\widehat c$ also changes the effective radiation energy reservoir. This matters when radiation contributes appreciably to the coupled thermal response. The static-medium equations further omit velocity-dependent radiation advection and work; they are not intended to describe dynamic diffusion, where trapped radiation is carried with the gas. This can have implications within rapidly moving accreting flows.

The dust prescription combines stellar absorption from the \citet{Draine03} model with thermal opacities from \citet{Semenov+03}. This permits source-dependent stellar absorption and temperature-dependent thermal opacities, including changes in grain material. A common dust abundance and temperature connect the two treatments, but they do not constitute a single grain population with one continuous opacity spectrum. Their relative absorption and emission efficiencies therefore remain a physical modeling uncertainty that can affect equilibrium temperatures and radiation forces. Moreover, evaluating the infrared absorption mean at the evolved color temperature is a coarse approximation, and can produce lower temperatures in highly optically thick regions. The present model does not evolve grain sizes, composition, or separate grain temperatures.

The chemical network is similarly targeted at thermal evolution, and detailed chemical abundances are reserved for future work. Hydrogen, helium, and molecular hydrogen evolve out of equilibrium, whereas heavy-element abundances and much of their cooling are prescribed. The model therefore does not predict time-dependent C, O, or CO abundances, and its hot-metal cooling assumes collisional ionization equilibrium. Applications to molecular-line emission, strongly irradiated metal ions, or rapidly cooling metal-rich gas would require a more detailed treatment. The fixed molecular ortho-to-para ratio is another assumption that can influence the heat capacity during cold collapse \citep{Boley+07,Vaytet+14}.

Finally, at the present fiducial resolutions, the outflow model is a subgrid prescription for the propagation of injected material and its dependence on resolution; the fiducial resolution does not resolve magnetocentrifugal launching. DISPATCH is capable of reaching the scales relevant to this, which carries an opportunity and a caution for future work: high-resolution tests at AU scales or smaller may launch jets self-consistently. The model is, therefore, limited to these lower resolutions: at very high resolutions, it may add injected energy to an effect that is no longer subgrid.

Because of these limitations, it is important to check the accretion scale, feedback parameters, radiation transport timescales, and numerical resolution for particular applications when using the framework to infer stellar masses or accretion lifetimes.

\section{Conclusions}
\label{sec:conclusions}

We have presented a framework extension for following individual stellar growth and feedback within DISPATCH, combining directional stellar irradiation, diffuse radiation transport, thermochemistry, and mechanical outflows. These processes operate within an adaptive mesh with patch-local timesteps, allowing collapsing gas and stellar environments to be followed alongside more slowly evolving cloud structures. We have validated it with tests of gravitational collapse, radiation-driven expansion, illumination and shadowing by multiple sources, and outflow propagation, establishing the behaviour of the coupled method in these complementary regimes.

The immediate scientific opportunity is to determine how feedback modifies gas accretion flows on large scales and its impact on the inertial inflow model. Following both the accreting stars and the surrounding flows allows stellar growth to respond to irradiation, shielding, and mechanical feedback from the evolving population. Further applications ranging from cloud-scale zoom-in simulations to studies of individual protostellar environments can then address similar underlying questions: how the interaction between gas supply and stellar feedback sets the masses of stars and the efficiency with which their parent clouds form them.

\begin{acknowledgements}

The authors thank Mike Grudi\'c for helpful discussions and assistance with the development of the methods presented here. W.L. and P.P. acknowledge support from the National Science Foundation through grant AST-2408023. P.P. acknowledges support from the grants PID2020-115892GB-I00 and CEX2019-000918-M funded by MCIN/AEI/10.13039/501100011033. The Tycho supercomputer hosted at the SCIENCE HPC center at the University of Copenhagen was used for supporting the development. T.H. acknowledges support from the Independent Research Fund Denmark through grant No. DFF 802100350B. W.L. acknowledges OpenAI's ChatGPT for Academic Researchers program for access to tools used to assist with code development, debugging, and testing.  

\end{acknowledgements}

\bibliographystyle{aa}
\bibliography{cosmo}

\begin{appendix}
\nolinenumbers
\section{Reaction Rates, Cooling Data, and Opacities}
\label{app:microphysics}

\subsection{Sources of the Microphysical Prescriptions}

Table~\ref{tab:microphysical-sources} summarizes the adopted reaction rates, cooling data, and dust opacities. Atomic losses are evaluated using the evolved ion fractions. Collisional ionization removes the corresponding threshold energy, while recombination cooling uses thermal-loss coefficients consistent with the adopted recombination prescription. Photoionized gas uses case-B radiative recombination under the on-the-spot approximation, with helium dielectronic recombination included separately.

\begin{table*}[t]
\centering
\caption{Principal sources of the thermochemical prescriptions.}
\label{tab:microphysical-sources}
\begin{tabular}{p{0.29\textwidth}p{0.65\textwidth}}
\toprule
Process & Prescription \\
\midrule
H/He electron-impact ionization & \citet{Voronov97} \\
H/He radiative recombination & Case A: \citet{VernerFerland96}; case B: \citet{HuiGnedin97} \\
Atomic collisional excitation & Ground-state CHIANTI data \citep{Dufresne+24} \\
Recombination cooling; free--free emission & \citet{HuiGnedin97,Rosdahl+13}; \citet{Katz+96} \\
H$_2$ grain formation and gas-phase reactions & \citet{GloverJappsen07,GloverAbel08}; associative detachment updated to \citet{Kreckel+10}, and H--H--H formation to \citet{Forrey13} \\
H$_2$ collisional cooling & Collider-dependent low-density and LTE prescriptions of \citet{GloverAbel08}, with the H$^+$ coefficient of \citet{Glover15} \\
Photoelectric heating and grain recombination & \citet{BakesTielens94,Wolfire+03} \\
Cosmic-ray heating & Atomic and molecular yields based on \citet{BialySternberg19,Glassgold+12}, with the reduced charge closure described below \\
Cold metal cooling & Fine-structure statistical equilibrium and an aggregate molecular term; see text \\
Hot metal cooling & Metal-only CIE data from \citet{Stofanova+21}, reweighted to Table~\ref{tab:metal-abundances} \\
Stellar dust absorption & $R_V=5.5$ carbonaceous--silicate model \citep{WeingartnerDraine01,Draine03} \\
Thermal dust opacity & Normal-composition composite aggregates \citep{Semenov+03} \\
\bottomrule
\end{tabular}
\end{table*}

\subsection{Molecular Reaction Heating and Shielding}

The gas heating associated with H$_2$ formation depends on the formation channel and on collisional de-excitation of the newly formed molecule. We adopt the following energies per formation event:
\begin{equation}
 \begin{split}
 Q_{\rm gr}&=(0.2+4.28q)\,\mathrm{eV},\\
 Q_{\rm H^-}&=3.53q\,\mathrm{eV},\qquad
 Q_{\rm 3b}=4.48\,\mathrm{eV},
 \end{split}
 \label{eq:app-h2-heat}
\end{equation}
where $q=n_{\rm H}/(n_{\rm H}+n_{\rm crit})$ is the collisional quenching fraction and $n_{\rm crit}$ depends on the local collider mixture \citep{GloverJappsen07}. The grain prescription assigns $0.2\,\mathrm{eV}$ to translational heating and at most $4.28\,\mathrm{eV}$ to quenched excitation. This is our adopted partition of the $4.48\,\mathrm{eV}$ binding energy, informed by \citet{GloverJappsen07,BialySternberg19}.
Collisional dissociation removes $4.48\,\mathrm{eV}$ per molecule; the proton charge-transfer channel removes $1.83\,\mathrm{eV}$. LW dissociation supplies $0.4\,\mathrm{eV}$ of fragment heat, with an additional effective $17q\,\mathrm{eV}$ per dissociation from associated UV pumping \citep{GloverJappsen07}. Formation and destruction contribute separately even when the molecular abundance is in equilibrium.

Grain formation uses the sticking and desorption prescription compiled by \citet{GloverJappsen07}, scaled by the surviving grain surface area. Three-body formation with H$_2$ as the third body uses one eighth of the H--H--H coefficient of \citet{Forrey13}. The H$_2$--H$^+$ cooling coefficient follows \citet{Glover15}; the other molecular cooling coefficients follow \citet{GloverAbel08}.

Stellar LW shielding is described in Section~\ref{sec:direct-radiation}. For the imposed ambient field, we instead estimate $N_{\rm H_2}=f_2N_{\rm H}/2$ from the local column and use the \citet{WolcottGreen+11} fit with exponent $1.1$ and thermal Doppler width. The two contributions are added after their respective shielding corrections.

\subsection{Cosmic Rays and Weakly Ionized Gas}

We prescribe the cosmic-ray ionization rate per atomic hydrogen atom as
\begin{equation}
 \zeta(N_{\rm H})=\zeta_0
 \left(1+\frac{N_{\rm H}}{N_{\rm CR}}\right)^{-\beta_{\rm CR}},
 \label{eq:app-cr-attenuation}
\end{equation}
with reference parameters $\zeta_0=2\times10^{-16}\,\mathrm{s^{-1}}$,
$N_{\rm CR}=10^{24}\,\mathrm{cm^{-2}}$, and
$\beta_{\rm CR}=0.4$. These specify an empirical attenuation
model. Heating uses composition-dependent atomic and molecular
yields based on \citet{BialySternberg19,Glassgold+12}; the
molecular chemical-heating contribution uses a reduced H$_3^+$
recombination branch.

In weakly ionized gas, unresolved molecular and metal ions contribute
to the electron abundance. Their equilibrium fractions are
\begin{align}
 x_{\rm m}&=\frac{\zeta f_2}
 {\alpha_{\rm m}n_{\rm H}x_{\rm e}+g_{\rm m}},\\
 x_{\rm M}&=\frac{\zeta_{\rm M}a_{\rm M}}
 {\zeta_{\rm M}+\alpha_{\rm M}n_{\rm H}x_{\rm e}+g_{\rm M}},
 \label{eq:app-charge-closure}
\end{align}
with charge neutrality giving
$x_{\rm e}=x_{\rm HII}+x_{\rm HeII}+2x_{\rm HeIII}+x_{\rm m}+x_{\rm M}$.
All fractions are relative to hydrogen nuclei.
Here $\zeta_{\rm M}=3.83\zeta$,
$a_{\rm M}=1.4\times10^{-4}Z'$, and
$\alpha_{\rm m}=4.5\times10^{-6}(T/\mathrm{K})^{-0.65}
\,\mathrm{cm^3\,s^{-1}}$.
Metal recombination follows the C-like fit compiled by
\citet{GloverJappsen07}. The grain-neutralization rates $g_i$
use geometric capture by $0.1\,\mu\mathrm{m}$ grains of material
density $2.4\,\mathrm{g\,cm^{-3}}$, scaled to the local dust abundance.
This approximation treats the ions as trace species in equilibrium
and does not evolve a grain-charge distribution.

Photoelectric heating and grain-recombination cooling follow
\citet{BakesTielens94,Wolfire+03}, with $\phi_{\rm PAH}=0.5$.
The small-grain photoelectric response and grain surface area can
be varied independently of the dust absorption opacity. The reference
ambient FUV intensity is $G_0=1.7$ Habing; external heating is
disabled in tests intended to isolate radiative cooling.

\subsection{Reduced Metal Cooling}

Cold metal cooling includes a two-level C$^+$-like coolant,
the three ground-term levels of neutral oxygen, and an aggregate
molecular contribution. Fine-structure populations include
collisions, the background radiation bath, and a local escape
probability. Reference cold-gas abundances are
$1.4\times10^{-4}Z'$ and $3.2\times10^{-4}Z'$ per H nucleus
for the C$^+$-like and oxygen coolants, respectively.
The aggregate molecular term uses the column-dependent density
saturation described by \citet{Hopkins+23}.

Warm ionized gas includes a nebular contribution with prescribed
ion fractions, following the approach of \citet{Kim+23}.
O, N, and Ne are divided $80:20$ between singly and doubly ionized
states, and S is divided equally; level populations depend on
temperature and electron density. Between $20000$ and
$35000\,\mathrm{K}$, the low-temperature metal contribution is
replaced smoothly by the hot CIE contribution:
\begin{align}
 \mathcal L_Z&=(1-w)\mathcal L_{Z,\rm low}
       +wZ'n_{\rm e}n_{\rm H}\Lambda_{Z,\rm CIE}(T),\\
 w&=s^2(3-2s),\qquad
 s=\frac{T-20000\,\mathrm{K}}{15000\,\mathrm{K}},
 \label{eq:app-metal-join}
\end{align}
with $w=0$ below the interval and $w=1$ above it.

The hot coefficient is assembled from the data in
\citet{Stofanova+21} using the abundances in
Table~\ref{tab:metal-abundances}, and excludes hydrogen and helium.
The cold-gas coolant abundances incorporate depletion, whereas
the hot table uses the listed reference mixture. A single $Z'$
scales each prescribed pattern; heavy-element ion fractions and
depletion are not evolved.

\begin{table}[t]
\caption{Reference abundances used for the hot metal cooling table.}\label{tab:metal-abundances}
\centering
\begin{tabular}{lc@{\qquad}lc}
\toprule
Element & $n_i/n_{\rm H}$ & Element & $n_i/n_{\rm H}$ \\
\midrule
C  & $2.95\times10^{-4}$ & N  & $7.41\times10^{-5}$ \\
O  & $5.37\times10^{-4}$ & Ne & $9.33\times10^{-5}$ \\
Na & $2.04\times10^{-6}$ & Mg & $4.37\times10^{-5}$ \\
Al & $2.95\times10^{-6}$ & Si & $3.55\times10^{-5}$ \\
S  & $1.45\times10^{-5}$ & Ar & $2.75\times10^{-6}$ \\
Ca & $2.14\times10^{-6}$ & Fe & $3.47\times10^{-5}$ \\
Ni & $1.74\times10^{-6}$ & & \\
\bottomrule
\end{tabular}

\end{table}
\end{appendix}

\end{document}